\documentclass[a4paper]{article}

\usepackage{jheppub} 
                     
\usepackage{graphicx,color}
 \usepackage{bm}
   \usepackage{amsmath}
    \usepackage{amssymb}    
     \usepackage{pifont}
      \usepackage{srcltx}
 \usepackage{stmaryrd}

\usepackage{standalone}
\usepackage{slashed}
\usepackage{multirow}
\usepackage{tabu}
\usepackage{hhline}
\usepackage{colortbl}

\newcommand{\Ds}{\displaystyle}

\newcommand{\nn}{\nonumber}

\newcommand{\Tr}{\mathrm{Tr}}

\renewcommand{\(}{\left(}
\renewcommand{\)}{\right)}
\renewcommand{\[}{\left[}
\renewcommand{\]}{\right]}

\renewcommand{\vec}[1]{\bm{#1}}
\newcommand{\fnot}[1]{\slashed{#1}}

\title{Kinematic power corrections in TMD factorization theorem}

\author{Alexey Vladimirov}

\affiliation{Departamento de F\'isica Te\'orica \& IPARCOS, Universidad Complutense de Madrid, E-28040 Madrid, Spain}

\emailAdd{alexeyvl@ucm.es}

\preprint{IPARCOS-UCM-23-046}

\abstract{
This work is dedicated to the study of power expansion in the transverse momentum dependent (TMD) factorization theorem. Each genuine term in this expansion gives rise to a series of kinematic power corrections (KPCs). All terms of this series exhibit the same properties as the leading term and share the same nonperturbative content. Among various power corrections, KPCs are especially important since they restore charge conservation and frame invariance, which are violated at a fixed power order. I derive and sum a series of KPCs associated with the leading-power term of the TMD factorization theorem. The resulting expression resembles a hadronic tensor computed with free massless quarks while still satisfying a proven factorization statement. Additionally, I provide an explicit check of this novel form of factorization theorem at the next-to-leading order (NLO) and demonstrate the restoration of the frame-invariant argument of the leading-power coefficient function. Numerical estimations show that incorporating the summed KPCs into the cross-section leads to an almost constant shift, which may help to explain the observed challenges in the TMD phenomenology.
}

\begin{document} 
\allowdisplaybreaks
\maketitle 

\section{Introduction}

The transverse momentum dependent (TMD) factorization theorem is known to be an invaluable tool in understanding the transverse momentum distribution of inclusive and semi-inclusive processes across a wide range of energy regimes. The transverse momentum distribution of Z-boson or Higgs bosons, semi-inclusive deep-inelastic scattering (SIDIS), single- and double-spin asymmetries -- these are only a few examples of successes of TMD factorization approach (for review see \cite{Angeles-Martinez:2015sea, Boussarie:2023izj}). As the list of accomplished applications continues to expand, it becomes evident that the current form of TMD factorization has practical limitations. It can only be confidently applied within a restricted phase-space, which leads to many problems and tensions already today (see f.i. discussions in \cite{Vladimirov:2019bfa, Bury:2021sue, Bacchetta:2022awv}), and will be critical for the future generation of colliders \cite{AbdulKhalek:2021gbh, Accardi:2023chb}. Therefore, a systematic extension and refinement of the TMD factorization approach are crucial priorities for modern studies.

A principal feature of the TMD factorization approach is its ability to systematically account for the transverse momenta of partons within the hadron. As a result, the main application domain is processes characterized by transverse momenta compatible with typical hadronic scales. At the same time, the characteristic longitudinal momenta must be large enough to justify the application of the parton model. In this study, I focus on the Drell-Yan reaction ($h_1+h_2\to \gamma^*+X$), which is the classical case of the TMD factorization \cite{Collins:1984kg}. In this particular reaction, the relevant momenta are the transverse ($q_T$) and longitudinal ($q^\pm$) components of the produced-photon momentum.

The leading term of the TMD factorization theorem is derived \cite{Collins:2011zzd, Becher:2010tm, Echevarria:2011epo} in the limit of the longitudinal momenta being greater than any other scales. In the position-space representation, the leading term has the general form (omitting the scaling arguments for brevity)
\begin{eqnarray}\label{into:1}
\frac{d\sigma}{dq_T}\sim \sigma_0\int_{-\infty}^{\infty} \frac{d^2b}{(2\pi)^2} e^{i(bq_T)}~C_{0}~ \widetilde{F}_1(x_1,b)\widetilde{F}_2(x_2,b),
\end{eqnarray}
where $\widetilde{F}_{1,2}$ are TMD parton distribution functions (TMDPDFs) in the Drell-Yan case, $x_{1,2}$ are the longitudinal momentum fractions of partons. $C_0$ is the hard coefficient function which depends on $q^+q^-/\mu^2$ with $\mu$ being the factorization scale. TMD distributions satisfy the pair of evolution equations \cite{Aybat:2011zv, Chiu:2012ir, Scimemi:2018xaf}, which have multiplicative kernels in the position space. Due to it, the position-space representation (\ref{into:1}) is convenient in practice. The momentum-space representation of (\ref{into:1}) is
\begin{eqnarray}\label{into:2}
\frac{d\sigma}{dq_T}\sim \sigma_0\int_{-\infty}^{\infty} d^2k_{1T} d^2 k_{2T} ~C_{0}~\delta^{(2)}(q_T-k_{1T}-k_{2T}) F_1(x_1,k_{1T}) F_2(x_2,k_{2T}),
\end{eqnarray}
where $F_{1,2}$ are TMDPDFs in the momentum space obtained by the Fourier transform of $\widetilde{F}_{1,2}$. Explicit examples of these structures for various processes can be found in refs.\cite{Mulders:1995dh, Bacchetta:2006tn, Arnold:2008kf}. The perturbative elements of the factorization formula (\ref{into:1}) (the hard coefficient function and the evolution kernels) are known up to next-to-next-to-next-to-next-to-leading order (N$^4$LO), making it one of the most developed formula in the perturbation theory.

The corrections to the leading term are suppressed by powers of $q^\pm$. To simplify the exposition, I generally refer to the large scale as $Q$ ($\sim q^+ \sim q^-$). The power corrections can be categorized into four conceptual types:

\textbf{$k_T/Q$ power corrections:} In the position-space representation, these corrections manifest as transverse derivatives of TMD distributions, while in the momentum representation, as the powers of $k_T$'s. These corrections are commonly known as \textit{kinematic power corrections} (KPCs).

\textbf{$q_T/Q$ power corrections:} In the position space, these corrections arise as terms accompanied by inverse powers of $b$. Once integrated, such factors transform into powers of $q_T$.

\textbf{$\Lambda/Q$ power corrections:} These corrections indicate terms that contain nonperturbative distributions of higher twist. Generally, a TMD distribution of twist-n is considered as a $(\Lambda/Q)^{n-2}$ correction. These corrections are often referred to as \textit{higher twist} power corrections.

\textbf{Target-mass corrections:} These corrections account for the finite mass of the hadron. Currently, they are the least studied corrections.

Let me emphasize the difference between $k_T/Q$ and $q_T/Q$ power corrections since they were not distinguished until recently. Indeed, in the momentum-space representation (\ref{into:2}), these corrections can be confused with each other due to the relation $q_T=k_{1T}+k_{2T}$. Nonetheless, they have conceptually different origins. The $q_T/Q$-corrections are related to the common ultra-violet (UV) singular behavior at $b\to 0$ and are merely a part of the hard scattering sub-process. The $k_T/Q$-corrections are UV-finite and are a part of low-energy dynamics. In the position space, the $k_T/Q$-corrections appear as the transverse derivatives of $\tilde F(x,b)$'s that turn to parton's $k_T$'s in the momentum space representation, which suggests the name. Therefore, these corrections are entirely independent from the theory point of view. Practically, these power corrections are also distinct. KPCs are non-vanishing in the limit $q_T\to 0$ (the traditional TMD limit) and remain of the same order for larger $q_T$. On the contrary, the $q_T/Q$-corrections vanish at $q_T\to0$ and grow at larger $q_T$ (as suggested by name), and constitute the major part of the famous Y-term \cite{Collins:1984kg, Collins:2011zzd, Boussarie:2023izj}. Below, I present more details on their definition.

Together, four types of power corrections form a complex system, which is discussed further in the remaining part of the introduction. Importantly, each type of power correction exhibits distinct theoretical and practical features, making some corrections more significant than others in different circumstances. In this study, I present the derivation, summation, and discussion of (pure) KPCs and examine their practical importance.

In recent years, there has been significant progress in the study of power corrections to the TMD factorization theorem \cite{Balitsky:2017gis, Balitsky:2020jzt, Inglis-Whalen:2021bea, Balitsky:2021fer, Vladimirov:2021hdn, Ebert:2021jhy, Rodini:2022wic, Rodini:2023plb}. This progress has been made possible by formulating the TMD factorization on the operator level, as opposed to the method-of-region analysis \cite{Collins:2011zzd} used in the original derivation of the leading power (LP) term. As a matter of fact, the power corrections are simpler to analyze with operator methods since they allow for unambiguous identification of the different types of power corrections. This is crucial for a comprehensive understanding of the structure of the factorization theorem beyond the LP approximation. Still, the status of the TMD factorization theorem beyond the LP approximation remains an open problem. Currently, there exists a complete expression for TMD factorization at the next-to-leading power (NLP) level. It has been derived with three different methods \cite{Balitsky:2021fer, Vladimirov:2021hdn, Ebert:2021jhy} (and agrees between them), and checked at next-to-leading order (NLO) \cite{Vladimirov:2021hdn, Rodini:2022wic, Rodini:2022wki}. Furthermore, several general statements (such as the all-order structure of rapidity divergences for twist-three distributions \cite{Vladimirov:2021hdn, Ebert:2021jhy}, the cancellation of special rapidity divergences \cite{Rodini:2022wic, Rodini:2022wki}) provide us hope that the NLP TMD factorization is valid at all perturbative orders. The research on the next-to-next-to-leading power (NNLP) TMD factorization is very limited \cite{Balitsky:2020jzt, Inglis-Whalen:2021bea, Balitsky:2021fer}. 

The straight evaluation of power corrections is impractical. Already at NNLP, one faces many novel theoretical problems and a vast number of structures. Therefore, to proceed further, it is crucial to understand and account for the general hierarchy of power-suppressed terms in the TMD factorization framework. This hierarchy can be deduced from the power counting and the dimension analysis. Naturally, there is a degree of arbitrariness in such a generalization. Nonetheless, if TMD factorization holds at higher powers, one should expect it to exhibit the following general structure (omitting target-mass corrections for simplicity)
\begin{eqnarray}\label{general-structure-1}
&&W^{\mu\nu}=\frac{1}{N_c}\int \frac{d^2b}{(2\pi)^2} \, e^{-i(q_Tb)}\Bigg\{
\\\nn && \phantom{+\frac{1}{Q\phantom{^1}}\Big(D^0}\Phi_\mathbf{2}\times \Phi_\mathbf{2}
\\\nn && +\frac{1}{Q\phantom{^1}}\Big(D\phantom{^1}\Phi_\mathbf{2}\times \Phi_\mathbf{2}+\phantom{D^0}\Phi_\mathbf{2}\times \Phi_\mathbf{3}\Big)
\\\nn && +\frac{1}{Q^2}\Big(D^2\Phi_\mathbf{2}\times \Phi_\mathbf{2}+D\phantom{^1}\Phi_\mathbf{2}\times \Phi_\mathbf{3}+\phantom{D}\Phi_\mathbf{3}\times \Phi_\mathbf{3}+\phantom{D}\Phi_\mathbf{2}\times \Phi_\mathbf{4}+\frac{\Phi_\mathbf{2}\times \Phi_\mathbf{2}}{b^2}\Big)
\\\nn && +\frac{1}{Q^3}\Big(D^3\Phi_\mathbf{2}\times \Phi_\mathbf{2}+D^2\Phi_\mathbf{2}\times \Phi_\mathbf{3}+D\Phi_\mathbf{3}\times \Phi_\mathbf{3}+D\Phi_\mathbf{2}\times \Phi_\mathbf{4}+\frac{D\Phi_\mathbf{2}\times \Phi_\mathbf{2}}{b^2}+\Phi_\mathbf{3}\times \Phi_\mathbf{4}+...\Big)
\\\nn &&+\cdots \Bigg\}^{\mu\nu},
\end{eqnarray}
where $\Phi_\mathbf{n}$ is a TMD distribution of twist-n, which depends on $b$ and some number of collinear-momentum fractions. The power of $D$ indicates the number of boost-invariant transverse derivatives (defined in (\ref{def:D})) in the term, which acts to TMD distributions in various combinations. The symbol $\times$ indicates an integral convolution in the collinear-momentum fractions. Each term is equipped with a coefficient function which is not shown. The first line in the brackets is the LP term (\ref{into:1}). The second line is the NLP term (see \cite{Balitsky:2020jzt, Vladimirov:2021hdn}), and so on. Each term in expression (\ref{general-structure-1}) represents a complicated composition of various distributions. 

In expression (\ref{general-structure-1}), one can easily recognize power corrections of different types. So, the first term in the third line ($D\Phi_\mathbf{2}\times \Phi_\mathbf{2}$) is the correction $\sim k_T/Q$, and the second term ($\Phi_\mathbf{2}\times \Phi_\mathbf{3}$) is the correction $\Lambda/Q$. In the fourth line, the first term ($D^2\Phi_\mathbf{2}\times \Phi_\mathbf{2}$) is $k_T^2/Q^2$, the second term ($D\Phi_\mathbf{2}\times \Phi_\mathbf{3}$) is $k_T\Lambda/Q^2$, the third and forth terms ($\Phi_\mathbf{3}\times \Phi_\mathbf{3}$ and $\Phi_\mathbf{2}\times \Phi_\mathbf{4}$) are $\Lambda^2/Q^2$, and the last term ($\Phi_\mathbf{2}\times \Phi_\mathbf{2}/b^2$) is $q_T^2/Q^2$. And so on. 

In eqn.~(\ref{general-structure-1}), the terms with different order of $k_T/Q$ corrections but the same order of other corrections are aligned into columns. These columns represent a series of KPCs to the first term. Each column incorporates unique nonperturbative content, as it contains a combination of TMD distributions that do not appear in other columns. Hence, each column can be considered as an independent contribution to the TMD factorization theorem. All global properties expected from the hadronic tensor, such as charge conservation and frame invariance, must hold for each column individually.

It is important to emphasize that the decomposition (\ref{general-structure-1}) is based on the assumption that TMD distributions of different twists (i.e., $\Phi_\mathbf{n}$ and $\Phi_{\mathbf{m}}$ for $n\neq m$) are completely independent functions. This assumption holds if these distributions do not mix under TMD evolution. The requirement for non-mixture is a fundamental prerequisite because otherwise, power corrections of different types would entangle due to evolution or by higher perturbative orders of coefficient functions, and the presentation form (\ref{general-structure-1}) would be impossible. Moreover, it would violate the universality property of nonperturbative distributions\footnote{
In principle, one could allow for the triangular mixture, such that higher-twist distributions depend on lower-twist. The presence of a mixture is inconvenient and makes the analysis of the factorization theorem difficult. In this case, one can always eliminate the mixture by a proper change of the basis.
}.
Therefore, the problem of systematization of power corrections in TMD factorization is equivalent to the problem of the definition of TMD-twist. This work adopts the definition of TMD twist introduced in ref. \cite{Vladimirov:2021hdn}, which constitutes a TMD generalization of the concept of \textit{geometrical twist} (aka "dimension minus spin") used for collinear operators \cite{Brandt:1970kg, Gross:1971wn}. This definition ensures, at least, the independence of non-mixture under ultraviolet renormalization.

Apart from the theoretical challenges posed by power-suppressed terms, there is also a practical concern regarding their relevance. It is evident that a plethora of TMD distributions appear in the power corrections, and the number of new functions grows faster than the number of observables. Already at NLP, one encounters 32 twist-three TMDPDFs \cite{Rodini:2022wki} (in addition to the 8 TMDPDFs at the leading power). Identifying and constraining all these functions from experimental data alone is practically impossible. In this regard, higher twist power corrections can be characterized as "bad" as they introduce unknown functions (although offering insights into new facets of quantum physics). The other types of power corrections are "better" as they do not increase the number of unknowns but instead refine and extend the LP term.

Thus, it is sensible to incorporate power corrections by types rather than by orders. First, this approach preserves the validity of the factorization theorem, as different types of power corrections are additive and do not interfere with each other. Secondly, each type of correction contributes to a specific kinematic region and could be distinguished phenomenologically. Neglecting power corrections put the requirements to the applicability range of the factorization theorem. Each power correction leads to its own limitation, and therefore, the LP TMD factorization is valid in the regime
\begin{eqnarray}
k_T\ll Q,\quad q_T\ll Q,\quad \Lambda \ll Q, \quad m\ll Q.
\end{eqnarray}
Including power corrections of a specific type softens the corresponding restriction, while summing the power corrections of a specific type eliminates the restriction (or replaces it with a strict inequality). At low $Q$, all types of power corrections are significant simultaneously. Notably, the $q_T/Q$ corrections are special, as their magnitude is controlled by the experimental kinematics. By including only $q_T/Q$ corrections, the cross-section can be described up to $q_T\sim Q$, while the other corrections remain negligible (for large $Q$). The remaining corrections follow a hierarchy of $k_T>m>\Lambda$ and are valid in the regime $q_T\ll Q$.

In this work, I study KPCs for the LP term. Within the expression (\ref{general-structure-1}), these power corrections are represented by the first column. Notably, these power corrections are of utmost importance within the conventional range of application for TMD factorization $q_T\ll Q$. Primarily, KPCs play a crucial role in restoring the electromagnetic (EM) gauge invariance (charge conservation) and frame invariance of the hadronic tensor, both of which are explicitly broken in the LP term. Consequently, KPCs can be viewed as an integral part of the LP factorization, as the absence of these corrections render the LP factorization inconsistent. Furthermore, unlike other scales such as $q_T$, $\Lambda$, or $m$, the scale $k_T$ is not fixed but rather an integration variable (\ref{into:2}). This introduces a self-contradiction, as the factorization formula includes values of $k_T$ that exceed $Q$, contradicting the initial assumption that $Q$ is the largest scale. Moreover, the average value of $k_T$ is of the order of the ultraviolet cut-off due to the large-$k_T$ asymptotic behavior of TMD distributions ($\sim k_T^{-2}$). An extra point in favor of KPCs is that the factorization theorem for them does not require a dedicated proof. KPCs inherit the factorization property from the LP term, for which the theorem is already proven. Considering these reasons, it is sufficient to include KPCs independently of other power corrections.

The paper is structured as follows. In sec.~\ref{sec:twist} I review the concept of the TMD-twist and present in detail the simplest (but the most useful) example of twist-decomposition for the quark operator (sec.~\ref{sec:twist-example}) and for a general bi-quark TMD distribution (sec.~\ref{sec:twist-bi-quark}). Sec. \ref{sec:KPC-order-by-order} is dedicated to the derivation of KPCs order-by-order in power expansion. Here, I start from the sketch of the operator derivation of the LP term (sec.~\ref{sec:TMD-at-LP}) and continue to this derivation to all-powers (sec.~\ref{sec:KPC-all-order}). In sec. \ref{sec:KPC-order-by-order}, I inspect the influence of loop corrections for KPCs and derive the argument of the coefficient function. The summation of the KPC series is done in sec.~\ref{sec:sumKPC}. To perform the summation, I first sum the series of twist-two terms for a general TMD correlator (sec.~\ref{sec:tw2-general}) and then insert it into the factorized expression (sec.~\ref{sec:W-summed}). In sec.~\ref{sec:structure-general-main}, I inspect the properties of the summed expression, using the example of a completely unpolarized contribution to hadron tensor of Drell-Yan. The subsections \ref{sec:EM-example} and \ref{sec:frame-example} are devoted to the EM gauge invariance and the frame invariance, correspondingly. Finally, the estimation of numerical impact is presented in sec.~\ref{sec:numerics}.

\section{Twist decomposition in TMD factorization}
\label{sec:twist}

The definition of TMD-twist is vital for systematisation of power corrections. Given this definition, it becomes possible to decompose operators of any dimension into irreducible components and express their matrix elements in terms of universal and independent nonperturbative TMD distributions. A suitable definition of TMD-twist was proposed in reference \cite{Vladimirov:2021hdn}. TMD-twist is determined by a pair of geometrical twists associated with the semi-compact operators that constitute any TMD operator. This approach strictly ensures non-mixture of operators with distinct TMD-twists under evolution with the ultraviolet (UV) renormalization scale $\mu$. Although the non-mixture under the rapidity evolution remains questionable, this property has been demonstrated in \cite{Vladimirov:2021hdn} for a wide range of TMD distributions (given by quasi-partonic operators), which encompasses all TMD distributions of twist-two and twist-three. Furthermore, the equivalence of rapidity evolution between TMD distributions of TMD-twist two and three was established in \cite{Ebert:2021jhy}. Therefore, this definition is satisfactory for the current objective, namely the determination of KPCs for the LP term, as they solely include TMD distributions of twist-two.

In this section, I provide a review of the general definition of TMD-twist and present the simplest yet non-trivial example of twist-decomposition. This example serves as the foundation for the subsequent computation of KPCs.

\subsection{TMD-twist}
\label{sec:TMD-twist}

At any power of TMD factorization theorem TMD distributions have the same general structure,
\begin{eqnarray}\label{def:Psi-gen}
\widetilde{\Phi}_{AB}(\{z\}_A,\{z\}_B;b)\sim 
\langle p,s|U_A(\{z\}_A;b)U_B(\{z\}_B;0)|p,s\rangle,
\end{eqnarray}
where $U$ is a light-cone operator, labels $A$ and $B$ indicate the quantum numbers of $U$, and $\{z\}_A$ is a set of coordinates $z$. The operator $U(\{z\}_A;b)$ is a T-ordered product of QCD fields positioned at $(z_in+b)$ for $z_i \in \{z\}_A$.  The fields in the operator are connected by light-like Wilson lines, which continue to the light-cone infinity. So, the operator $U$ spans an infinite range, and for that reason it is called \textit{semi-compact}.

The renormalization of TMD distribution consists of two renormalization factors -- one for each operator $U$. These factors are independent since the distance between operators is transverse and thus UV finite. The evolution equation reads
\begin{eqnarray}\label{evol:A+B}
\mu^2 \frac{d}{d\mu^2}\widetilde{\Phi}_{AB}(\{z\}_A,\{z\}_B;b)
=
\(\tilde \gamma_A(\{z\}_A) + \tilde \gamma_B(\{z\}_B)\)\otimes
\widetilde{\Phi}_{AB}(\{z\}_A,\{z\}_B;b),
\end{eqnarray}
where $\tilde \gamma_A$ is the anomalous dimension of $U_A$ and $\otimes$ is the integral convolution in corresponding positions $\{z_i\}$. Note, that anomalous dimensions $\tilde \gamma$ contain the double logarithmic part, which is resulted from gluon propagated along Wilson line to the infinity (collinear singularity). The same singularity is presented in the rapidity renormalization factor (or soft factor). Their cancellation leads to a logarithm contribution $\sim \ln(\mu^2/\zeta)$ in the anomalous dimensions.

The equation (\ref{evol:A+B}) describes only the UV evolution, while TMD distributions also obey rapidity evolution equation (\ref{evol:zeta}). Currently, there is no general consensus regarding the rapidity divergences of operators that arise at higher powers. The only established result is that all quasi-partonic TMD operators conform to the leading power (LP) rapidity evolution \cite{Vladimirov:2021hdn}. Thus, it is plausible that the present definition of TMD-twist is incomplete, and some operators could potentially mix due to the rapidity evolution. However, this hypothetical situation would only apply to non-quasi-partonic TMD distributions, which have a minimum twist-four and emerge at NNLP for the first time, and thus, are irrelevant for the present discussion. 

The light-cone operators $U_A$ and $U_B$ have independent anomalous dimensions if they have different geometrical twist (defined as ``dimension-minus-spin'' of the operator). Consequently, TMD distributions $\widetilde{\Phi}_{AB}$ and $\widetilde{\Phi}_{A'B'}$ do not mix with each other if the geometrical twist of $U_A$ or $U_B$ is different from the geometrical twist of $U_{A'}$ or $U_{B'}$. It gives raise to a natural definition of TMD-twist for $\widetilde{\Phi}_{AB}$, as a pair of integer numbers $(N,M)$, where $N$($M$) is a geometrical twist of semi-compact operator $U_{A(B)}$.

Formally, the definition of geometrical twist is defined only for local operators. Its generalization for semi-compact operator can be done by the following procedure. First, one selects $L$ such that $\{|z|\}<|L|$, and drop the part of Wilson line from $Ln$ to infinity. Next, the operator is expanded as series at $L$. It takes a form $U_A(z,0)\sim \sum_n (z-L)^nU_{A,n}$, where $U_{A,n}$ are local operators.  The geometric twist of local operators $U_{A,n}$ is the geometric twist of semi-compact operator $U_A$. After the decomposition of operator into components with the same twist, each component can be summed over $n$. Finally, the limit $L\to \infty$ is taken. This method properly reconstructs the properties of semi-compact operators, as it is demonstrated in ref. \cite{Moos:2020wvd}.

An indication of twist for TMD operator by a single number (used so far), refers to $(N+M)$. For example, it was used in the expression (\ref{general-structure-1}), where $\Phi_{\mathbf{n}}=\Phi_{MN}$ with $N+M=n$. Meanwhile, all TMD distributions of twist-two have TMD-twist $(1,1)$. However, the single-number terminology is ambiguous beyond the twist-two case. For example, a TMD distribution of twist-three can have TMD-twist $(1,2)$ or $(2,1)$, which are two entirely independent distributions with separate evolution equations (see ref.\cite{Rodini:2022wki}). TMD distribution of twist-four can have TMD-twist $(1,3)$, $(3,1)$ or $(2,2)$, and so on. Still, the single-number indication is shorter and it is convenient to use if it does not create a confusion.

\subsection{Example of twist-decomposition}
\label{sec:twist-example}

As an example of twist decomposition let me consider the simplest semi-compact operator that appears in the TMD factorization. It is a quark field with an attached semi-infinite Wilson line
\begin{eqnarray}\label{example:U1}
U_{q}(0,0_T)=[-\infty n,0]q(0).
\end{eqnarray}
where $[an,bn]$ is a Wilson line along $n$
\begin{eqnarray}
[a n,bn]=P\exp\(-ig \int_{a}^b ds A_+(sn)\).
\end{eqnarray}
As usually, the vectors $n^\mu$ and $\bar n^\mu$ are two independent light-cone vectors with normalization
\begin{eqnarray}\label{def:n-barn}
n^2=\bar n^2=0,\qquad (n\bar n)=1.
\end{eqnarray}
The decomposition of any vector $v^\mu$ reads
\begin{eqnarray}
v^\mu=\bar n^\mu v^++n^\mu v^-+v_T^\mu,
\end{eqnarray}
where $v^+=(nv)$, $v^-=(\bar nv)$ and $v_T$ is the transverse component orthogonal to $(n,\bar n)$-plane $v_T^{\mu}=g_T^{\mu\nu}v_\nu$ with
\begin{eqnarray}
g_T^{\mu\nu}=g^{\mu\nu}-n^\mu \bar n^\nu-\bar n^\mu n^\nu.
\end{eqnarray}

Following the procedure described above the operator (\ref{example:U1}) can be presented as
\begin{eqnarray}\label{example:1}
U_{q}(0,0_T)=\lim_{L\to-\infty}\sum_{n=0}^\infty \frac{i^nL^n}{n!}D_+^nq(L n),
\end{eqnarray}
where $D_\mu=\partial_\mu-igA_\mu$ is the QCD covariant derivative. The local operator $D_{\mu_1}...D_{\mu_n}q_i$ has the mass-dimension $n+3/2$. It is a Lorenz tensor of mixed nature with $n$ vector indices and one spinor index. The maximum possible spin of this tensor is $n+1/2$. To achieve it, one should symmetrize indices $\mu$, subtract traces (in the present case, it automatically achieved by contraction with $n^{\mu_1}...n^{\mu_n}$), and make the spinor index $\gamma$-traceless \cite{Rarita:1941mf}, i.e. such that $\gamma^{\mu_j}D_{\mu_1}...D_{\mu_j}...D_{\mu_n}q=0$ (for a more formal discussion see refs.\cite{DelgadoAcosta:2015ypa, Geyer:1999uq}). For the present tensor (that is contracted with $n^{\mu_1}...n^{\mu_n}$) it implies that the maximum-spin operator should vanish after multiplication by $\gamma^-$. It leads to a natural decomposition of the spinor into ``good'' and ``bad'' components \cite{Jaffe:1991kp}, as $q=\xi_{\bar n}+\eta_{\bar n}$, where
\begin{eqnarray}\label{def:good-bad-components}
\xi_{\bar n}=\frac{\gamma^-\gamma^+}{2}q, \qquad \eta_{\bar n}=\frac{\gamma^+\gamma^-}{2}q.
\end{eqnarray}
Here, the subscript $\bar n$ indicates that these components are defined with respect to $\gamma^-$. Since $\gamma^-\xi=0$, the operator $D_+^n\xi$ has the maximum spin, and its geometrical twist equals one. Summing the series over $n$ and limiting $L\to-\infty$, the twist-one part of (\ref{example:U1}) is
\begin{eqnarray}\label{def:U1}
U_{q}(0,0_T)\Big|_{\text{tw-1}}=U_{1}(0,0_T)=[-\infty n,0]\xi_{\bar n}(0).
\end{eqnarray}

The remaining part of $U_{q}$ is $[-\infty n,0]\eta_{\bar n}(0)$. It is twist-two, but the field $\eta$ is not dynamically independent. The components $\eta$ and $\xi$ are related to each other via the QCD equation of motion (EOM). The EOM is $\fnot D q=0$ for massless quark. After projection (\ref{def:good-bad-components}) EOM splits into two equations, which are convenient to write as
\begin{eqnarray}\label{def:EOMs}
D_+\eta_{\bar n}=-\frac{1}{2}\gamma^+\fnot D_T\xi_{\bar n},\qquad D_-\xi_{\bar n}=-\frac{1}{2}\gamma^-\fnot D_T\eta_{\bar n}.
\end{eqnarray}
Inserting the left equation into eqn. (\ref{example:1}) and commuting the transverse derivative to the outer position one gets
\begin{eqnarray}
D_+^n\eta_{\bar n}(L n)&=&
\frac{-1}{2}\gamma^+\gamma^\mu_T D_+^{n-1}D_\mu\xi_{\bar n}(L n)
\\\nn 
&=&
\frac{-1}{2}\gamma^+\gamma^\mu_T\(
D_\mu D_+^{n-1}\xi_{\bar n}(L n)
+ig\sum_{m=0}^{n-1} D_+^{n-m} F_{\mu+} D_+^m\xi_{\bar n}(L n)\),
\end{eqnarray}
where $F_{\mu\nu}=ig^{-1}[D_\mu,D_\nu]$ is the gluon field-strength tensor. The subscript on a gamma-matrix indicates that its index is transverse. The summing over $n$ gives
\begin{eqnarray}\label{example-bad-comp}
[L n,0]\eta_{\bar n}(0)&=&
\frac{-1}{2}\gamma^+\gamma^\mu_T \int_{-\infty}^0 ds
\Big(D_\mu[Ln,sn]\xi_{\bar n}(sn)
+ig \int_{L}^s ds_1
[Ln,s_1n]F_{\mu+}[s_1n,sn]\xi_{\bar n}(sn)\Big).
\end{eqnarray}
The spin of the last term is $n-1/2$, since it is traceless and $\gamma$-traceless, and has all, except one, indices symmetrized. Thus, the last term is twist-two operator.

The final expression for twist-decomposition of $U_q$ can be assembled. Let the transverse component of the gluon field vanish at light-cone infinity\footnote{
This assumption is valid in any regular gauge, and could be used as the boundary condition of singular gauges (such as the light-cone gauge). It is known that in the absence of such condition one should attach the link along transverse direction $[-\infty n+\infty_T,-\infty n]$ \cite{Belitsky:2002sm}. However, such link could not be incorporated into the definition of twist in any form. Therefore, all presented consideration is done assuming that component of the gluon field is vanishing at appropriate light-cone infinity. Since the final result is gauge invariant the necessary transverse links can be inserted into the formulas after all derivatives are resolved.
}, so $\lim_{L\to \infty}D_\mu(L)=\partial_\mu$. Then combining (\ref{def:U1}) and (\ref{example-bad-comp}), one receives
\begin{eqnarray}\label{example:2}
U_q(0,0_T)=U_1(0,0_T)-\frac{\gamma^+\gamma^\mu_T}{2}\int_{-\infty}^0 ds \partial_{\mu}U_1(s,0_T)
-i\frac{\gamma^+\gamma^\mu_T}{2}\int_{-\infty}^0 ds \int_{-\infty}^{s} ds_1 U_{2,\mu}(s_1,s,0_T),
\end{eqnarray}
where $U_1$ is the twist-one operator (\ref{def:U1}) and $U_2$ is the twist-two operator
\begin{eqnarray}
U_{2,\mu}(z_1,z_2,0_T)=g [-\infty n,z_1n]F_{\mu+}[z_1n,z_2n]\xi_{\bar n}(z_2n).
\end{eqnarray}
Therefore, the operator $U_q$ is decomposed into operators of twist-one (first term) and twist-two (last term), and the \textit{total derivative} of twist-one operator (second term). The procedure presented above is standard for analysis of higher-twist collinear operators, see e.g. examples in refs. \cite{Jaffe:1996zw, Balitsky:1987bk, Belitsky:2000vx}. The only difference in the present case is the open spinor index, which should be extra treated in the tensor decomposition.

Note that the ``good'' and ``bad'' components has also different counting with respect to $\lambda\sim k_T/k_+$. Assuming that a quark field has large $k^+$ and almost massless $k^2\sim 0$, one finds that $\eta_{\bar n} \sim \lambda \xi_{\bar n}$ from EOMs (\ref{def:EOMs}). This counting is convenient to use for sorting operators in the power series, and it is widely used. 

\subsection{Twist-decomposition for bi-quark TMD distribution}
\label{sec:twist-bi-quark}

The procedure of twist-decomposition for semi-compact operators leads to the decomposition of TMD matrix elements to definite twist parts. Let me demonstrate it for the bi-quark TMD matrix element, which is also important for derivation of KPCs. The matrix element reads
\begin{eqnarray}\label{example:PHI}
\widetilde{\Phi}_{\bar qq}^{[\gamma^\mu]}(x,b)&=&
\int_{-\infty}^\infty \frac{dz}{2\pi}e^{-ixzp^+}
\langle p|\overline{U}_q(zn,b)\frac{\gamma^\mu}{2} U_q(0,0_T)|p\rangle,
\\\nn &=&
\int_{-\infty}^\infty \frac{dz}{2\pi}e^{-ixzp^+}
\langle p|\bar q(zn+b)[zn+b,-\infty n+b]\frac{\gamma^\mu}{2}[-\infty n,0]q(0)|p\rangle,
\end{eqnarray}
where $|p\rangle$ is a hadron state with momentum $p$, $b$ is a transverse vector and $\overline{U}=U^\dagger \gamma^0$. Being contracted with $n^\mu$ this matrix element results into the unpolarized and Sivers TMD distributions that are twist-two TMD distribution. For a general $\gamma^\mu$, the matrix element $\widetilde{\Phi}_{\bar qq}^{[\gamma^\mu]}$ does not represent a conventional distribution, but is a sum of distributions with different properties. It can be checked by computing the renormalization of matrix element (\ref{example:PHI}) and observing that it could not be expressed via $\widetilde{\Phi}^{[\gamma^\mu]}$ (see f.i. computations in refs.\cite{Rodini:2022wki, Gamberg:2022lju}). The reason is that the operator (\ref{example:PHI}) has an undefinite twist.

Inserting the decomposition (\ref{example:2}) into (\ref{example:PHI}), the correlator $\widetilde{\Phi}^{[\gamma^\mu]}_{\bar q q}$ turns into
\begin{eqnarray}\label{example:3}
&&\widetilde{\Phi}_{\bar qq}^{[\gamma^\mu]}(x,b)=
\int_{-\infty}^\infty \frac{dz}{2\pi}e^{-ixzp^+}\Bigg[
\langle p|\overline{U}_{1}(z,b)\frac{\gamma^\mu}{2}U_1(0,0_T)|p\rangle
\\\nn &&\quad
-\frac{1}{2}\int_{-\infty}^0 ds
\Big\langle p\Big|\overline{U}_{1}(z,b)\frac{
\gamma^\mu\gamma^+\gamma^\nu_T}{2}
\overrightarrow{\partial}_\nu 
U_1(s,0_T)
+
\overline{U}_{1}(z+s,b)\overleftarrow{\partial}_\nu\frac{\gamma_T^\nu\gamma^+\gamma^\mu 
}{2}U_1(0,0_T)\Big|p\Big\rangle
\\\nn &&
\quad
+\frac{1}{4}\int_{-\infty}^0 ds dt 
\langle p|\overline{U}_{1}(z+s,b)\overleftarrow{\partial}_\nu
\frac{\gamma_T^\nu\gamma^+\gamma^\mu \gamma^+\gamma^\rho_T}{2}
\overrightarrow{\partial}_\rho U_1(t,0_T)|p\rangle
\\\nn &&
\quad
+\frac{i}{2}
\int_{-\infty}^0 ds\int_{-\infty}^s ds_1
\\\nn && \qquad
\Big\langle p\Big|
\overline{U}_{2,\nu}(z+s,z+s_1,b)
\frac{\gamma^\nu_T\gamma^+\gamma^\mu}{2}U_1(0,0_T)
-
\overline{U}_{1}(z,b)
\frac{\gamma^\mu\gamma^+\gamma^\nu_T}{2}U_{2,\nu}(s_1,s,0_T)
\Big|p\Big\rangle
\\\nn &&\quad
-\frac{i}{4}
\int_{-\infty}^0 dt\int_{-\infty}^0 ds\int_{-\infty}^s ds_1
\Big\langle p\Big|
\overline{U}_{2,\nu}(z+s,z+s_1,b)
\frac{\gamma^\nu_T\gamma^+\gamma^\mu\gamma^+\gamma_T^\rho}{2}\partial_\rho U_1(t,0_T)
\\\nn && \qquad\qquad\qquad\qquad\qquad\qquad\qquad
-
\overline{U}_{1}(z+t,b)\overline{\partial}_\rho
\frac{\gamma^\rho_T\gamma^+\gamma^\mu\gamma^+\gamma^\nu_T}{2}U_{2,\nu}(s_1,s,0_T)
\Big|p\Big\rangle
\\\nn &&\quad
+\frac{1}{4}\int_{-\infty}^0 dt\int_{-\infty}^t dt_1\int_{-\infty}^0 ds\int_{-\infty}^s ds_1
\Big\langle p\Big|
\overline{U}_{2,\nu}(z+s,z+s_1,b)
\frac{\gamma^\nu_T\gamma^+\gamma^\mu\gamma^+\gamma_T^\rho}{2}U_{2,\rho}(t,t_1,0_T)\Big|p\Big\rangle
\Bigg]~,
\end{eqnarray}
where $\overline{U}=U^\dagger \gamma^0$. The matrix element in the first three lines are the TMD distributions of TMD-twist-(1,1). The lines fifth to seventh are TMD distributions with TMD-twists-(2,1) and (1,2). The last line is the TMD distribution of TMD-twist-(2,2). 


The gamma-structure of (\ref{example:3}) can be simplified using that the spinor indices of operators $U$ projected by matrices (\ref{def:good-bad-components}). For example, one finds that $\overline{U}_1\gamma^\mu U_1=\bar n^\mu \overline{U}_1\gamma^+ U_1$. The total derivatives of operators $U$ can be pushed outside of matrix elements using that
\begin{eqnarray}\label{tot-derivative=0}
\langle p|\partial_\mu (...)|p\rangle =0.
\end{eqnarray} 
After these simplifications the result reads
\begin{eqnarray}\label{example:4}
\widetilde{\Phi}^{[\gamma^\mu]}_{\bar q q}(x,b)&=&
\(\bar n^\mu+\frac{i}{xp^+}\frac{\partial}{\partial b_\mu}+\frac{n^\mu}{2(xp^+)^2}\frac{\partial^2}{\partial b^\nu \partial b_\nu}\)\widetilde{\Phi}^{[\gamma^+]}_{11}(x,b)+...~,
\end{eqnarray}
where $\Phi^{[\gamma^+]}_{11}$ is a TMD distribution of TMD-twist-(1,1)
\begin{eqnarray}\label{example:PHI11}
\widetilde{\Phi}^{[\Gamma]}_{11}(x,b)&=&
\int_{-\infty}^\infty \frac{dz}{2\pi}e^{-ixzp^+}
\langle p|\overline{U}_1(z,b)\frac{\Gamma}{2}U_1(0,0_T)|p\rangle,
\end{eqnarray}
and dots denote contributions with TMD-twist-(1,2), (2,1) and (2,2) resulted from the last five lines of (\ref{example:3}). The expression (\ref{example:4}) is the twist-two part of the correlator (\ref{example:PHI}).

The momentum space representation of TMD distribution is obtained by
\begin{eqnarray}\label{def:TMD-momentum}
\widetilde{\Phi}^{[\Gamma]}(x,b)=\int d^2k_T e^{i(bk_T)}\Phi^{[\Gamma]}(x,k_T).
\end{eqnarray}
Here and in the following, I use the (un)tilded notation to indicate the distributions in the (momentum) position space. The momentum space representation of (\ref{example:4}) is
\begin{eqnarray}\label{example:5}
\Phi_{\bar qq}^{[\gamma^\mu]}(x,k_T)&=&
\(\bar n^\mu+\frac{k_T^\mu}{xp^+}-\frac{n^\mu k_T^2}{2(xp^+)^2}\)\Phi^{[\gamma^+]}_{11}(x,k_T)+...~.
\end{eqnarray}
The derivatives over $b$ turn into powers of $k_T$.

It is also critical to observe that total derivatives of semi-compact operators does not always result to the definite-twist TMD distributions. It works out for total derivatives $\partial_+$ (which produces factors of $x$ in the momentum space), and $\partial_{\mu T}$ (which produces factors of $k_T^\mu$). However, total derivatives $\partial_-$ cannot be moved outside of the operator. Nonetheless, they could be always expressed via $\partial_+$, $\partial_{\mu T}$ and extra gluon fields, thanks to EOMs (\ref{def:EOMs}). The example is given in sec.~\ref{sec:KPC-all-order}.

The expression (\ref{example:5}) can be written as
\begin{eqnarray}\label{example:6}
\Phi^{[\gamma^\mu]}_{\bar q q}(x,k_T)&=&
2\int_{-\infty}^\infty dk^-\,  k^\mu \delta(k^2)\Phi^{[\gamma^+]}_{11}(x,k_T)+...~,
\end{eqnarray}
with $k^\mu=xp_+\bar n^\mu +k_T^\mu+n^\mu k^-$ being the four-momentum of the parton. This respresentation gives rise to an interpretation of the leading twist distributions, as the part of the hadron's structure carried by the free-quark approximation. In the sec.~\ref{sec:tw2-general} I provide a further generalization of eqn.(\ref{example:6}).

The twist-decomposition is a standard procedure for consideration of power corrections for factorization theorems, and the expressions (\ref{example:2}) and (\ref{example:4}) are rather typical for such kind of calculus. Many analogies can be found in the literature related to the collinear factorization. For example, the expressions (\ref{example:4}) and (\ref{example:5}) are ideologically analogous to the Wandzura-Wilczek approximation for collinear distributions \cite{Wandzura:1977qf}. In the present work, the main interest is the terms with total derivatives of the leading twist operator. These terms are responsible for generation of KPCs. It is analogous to the case of Deeply-Virtual Compton Scattering (DVCS) and Generalized parton distributions (GPDs) \cite{Braun:2011dg, Braun:2011zr}. 

Note that for the operator with a differently oriented light-like Wilson line, the  vectors $n$ and $\bar n$ should be swapped. For example, for decomposition for the operator $[-\infty \bar n,0]q(0)$ is the same as (\ref{example:2}) but with $n\leftrightarrow \bar n$. The ``good'' and ``bad'' components are defined as
\begin{eqnarray}\label{def:good-bad-components-n}
\xi_{n}=\frac{\gamma^+\gamma^-}{2}q, \qquad \eta_{n}=\frac{\gamma^-\gamma^+}{2}q.
\end{eqnarray}
The direction of Wilson line (future vs. past) does not play a role in this example.

For the following discussion, it is important to specify the evolution equation for twist-two TMDPDFs. these equations are \cite{Aybat:2011zv, Chiu:2012ir}
\begin{eqnarray}\label{evol:mu}
\mu^2\frac{d}{d\mu^2}\widetilde{\Phi}_{11}^{[\Gamma]}(x,b;\mu,\zeta)&=&\(\frac{\Gamma_{\text{cusp}}(\mu)}{2}\ln\(\frac{\mu^2}{\zeta}\)-\frac{\gamma_V(\mu)}{2}\)\widetilde{\Phi}_{11}^{[\Gamma]}(x,b;\mu,\zeta),
\\\label{evol:zeta}
\zeta\frac{d}{d\zeta}\widetilde{\Phi}_{11}^{[\Gamma]}(x,b;\mu,\zeta)&=&-\mathcal{D}(b,\mu)\widetilde{\Phi}_{11}^{[\Gamma]}(x,b;\mu,\zeta),
\end{eqnarray}
where $\Gamma_{\text{cusp}}$, $\gamma_V$ and $\mathcal{D}$ are the cusp, quark-vector and rapidity anomalous dimensions, correspondingly. The rapidity anomalous dimension is also known as the Collins-Soper kernel \cite{Collins:1981uk}. It is a nonperturbative function that is associated with the properties of QCD vacuum \cite{Vladimirov:2020umg}. In the momentum space the equation (\ref{evol:zeta}) turns to an integral equation \cite{Ebert:2016gcn}.

\section{KPCs for LP term}
\label{sec:KPC-order-by-order}

TMD operators of twist-two have a crucial property of incorporating only \textit{two} ``good'' components of fields (quark and anti-quark, or two gluon fields). This simplifies the procedure of their extraction from higher-dimensional operators and allows for the unambiguous elimination of higher twist contributions at an early stage of computation. As a result, the determination of KPCs for the LP term can be achieved at all orders of the power series, as presented in this section. For concreteness, I focus on the Drell-Yan process
\begin{eqnarray}\label{def:DY-reaction}
h_1(p_1) + h_2(p_2) \rightarrow \gamma^*(q) + X,
\end{eqnarray}
where the arguments denote the momenta of the respective particles. The generalization for other processes, such Semi-Inclusive Deep-Inelastic Scattering (SIDIS) or $e^+e^-\to h_1h_2+X$, is straightforward. I start with the schematic review of the derivation of LP term, and then generalize this example including KPCs.

\subsection{TMD factorization at LP}
\label{sec:TMD-at-LP}

\begin{figure}[t]
\begin{center}
\includegraphics[width=0.99\textwidth]{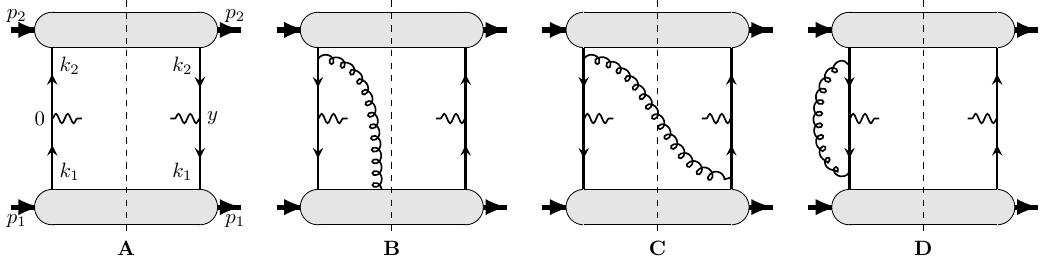}
\caption{\label{fig:diags}
Example of diagrams contributing to the DY process. Gray blobs show the hadron nonperturbative state. The vertical dashed line denote the insertion of complete set of states.
}
\end{center}
\end{figure}

The derivation of the  LP term of the TMD factorization theorem has been extensively discussed in numerous references employing various formulations. Here are only some of them \cite{Mulders:1995dh, Boer:2003cm, Bacchetta:2006tn, Arnold:2008kf, Becher:2010tm, Collins:2011zzd,  Echevarria:2011epo,  Vladimirov:2021hdn}. In this section, I provide an overview of the derivation from the perspective of operator manipulations, similar to the TMD operator expansion, Soft-Collinear effective theories or High-Energy expansion. Emphasis is done on the technical aspects that are relevant for further explanations. I omit detailed proofs that can be found in the literature.

The QCD part of the cross-section for a hard reactions is given by the hadronic tensor. For the Drell-Yan process (\ref{def:DY-reaction}) the hadronic tensor reads
\begin{eqnarray}\label{def:W}
W^{\mu\nu}=\int \frac{d^4 y}{(2\pi)^4}e^{-i(yq)}\sum_{X} \langle p_1,p_2|J^\mu(y)|X\rangle\langle X|J^\nu(0)|p_1,p_2\rangle,
\end{eqnarray}
where $J^\mu$ is the electromagnetic (EM) current
\begin{eqnarray}
J^\mu(y)=\bar q(y)\gamma^\mu q(y),
\end{eqnarray}
with $q$ being the quark field. The quark charges and flavor part of the current are omitted for brevity and restored in the following sections.

The kinematics of the process is defined by the photon momentum $q^\mu$ and the hadrons momenta $p_1^\mu$ and $p_2^\mu$. To avoid complications related to the target mass corrections, the hadrons are considered as massless $p_1^2=p_2^2=0$. These momenta introduce the natural system of vectors $n$ and $\bar n$ (\ref{def:n-barn})
\begin{eqnarray}\label{def:p1p2}
p_1^\mu=\bar n^\mu p_1^+,\qquad p_2^\mu=n^\mu p_2^-.
\end{eqnarray}

The leading perturbative order contribution to the hadronic tensor of Drell-Yan process (\ref{def:W}) is shown in fig.\ref{fig:diags}(A). This diagram and its charge-conjugated are convenient to write as
\begin{eqnarray}\label{KPC:1}
W^{\mu\nu}&=&\int \frac{d^4y}{(2\pi)^4}e^{-i(yq)}
\langle p_1|\langle p_2|J^\mu_{\bar nn}(y) J^\nu_{n\bar n}(0)+J^\mu_{n\bar n}(y) J^\nu_{\bar nn}(0)|p_1\rangle |p_2\rangle+...,
\end{eqnarray}
where dots represent the sub-leading terms, and
\begin{eqnarray}\label{KPC:J}
J^\mu_{\bar n n}=\bar q_{\bar n}\gamma^\mu q_n, \qquad J^\mu_{n \bar n}=\bar q_{n}\gamma^\mu q_{\bar n}.
\end{eqnarray}
The fields $q_{\bar n}$ ($q_n$) are created by $|p_1\rangle$ ($|p_2\rangle$) states, and, according to the parton model, have momentum almost along $p_1$($p_2$). Usually, fields $q_{\bar n}$ and $q_n$ are referred as collinear and anti-collinear fields. At LP approximation the states $|p_1\rangle$ and $|p_2\rangle$ do not interact with each other directly. The complete set of states in-between currents is omitted for simplicity.

To proceed further, one should specify the counting rules for the elements of equation (\ref{KPC:1}). For the TMD factorization the counting rules are  \cite{Becher:2010tm, Echevarria:2011epo, Vladimirov:2021hdn}
\begin{eqnarray}\label{counting-rules}
\partial_\mu q_{\bar n}\sim A_{\mu,\bar n}\sim \{1,\lambda^2,\lambda\}
,\quad
\partial_\mu q_{n}\sim A_{\mu,n}\sim \{\lambda^2,1,\lambda\},
\end{eqnarray}
where $\lambda \sim Q^{-1}$. The components are presented in the order $\{+,-,T\}$. In the regime $q^\pm \sim Q$ and $q_T\sim Q^0$ the vector $y$ satisfies the counting
\begin{eqnarray}\label{counting-rules-y}
y^\mu\sim Q^{-1}\{1,1,\lambda^{-1}\}
\end{eqnarray}
The inhomogeneity of counting for components of $y^\mu$ is the only conceptual difference between collinear and TMD factorizations. 

The counting rules imply that the interactions between fields along certain directions are power-suppressed. The field must be expanded around these directions\footnote{
For definiteness, the expansion is made at the origin. However, any other point can be used. The result is independent on this choice, thanks to eqn.(\ref{tot-derivative=0}). However, in the factorization theorems for off-forward processes, which involve generalized TMD (GTMD) distributions, such as \cite{Bhattacharya:2018lgm, Echevarria:2022ztg}, the result depends on the reference point. The independence on reference point should be restored in the sum of all KPCs similarly to the DVCS case \cite{Braun:2014paa}.
}
\begin{equation}\label{KPC:2}
q_{\bar n}(y)=q_{\bar n}(y^-n+y_T)+y^+\partial_-q_{\bar n}(y^-n+y_T)+...~,
\end{equation}
where the first term is $\sim \lambda^0$, the second term is $\sim \lambda^2$, and the dots indicate the further suppressed terms. Importantly, there is no expansion in the transverse direction, because $y^\mu_T\partial_{\mu }q_{\bar n}\sim \lambda^0$ according to (\ref{counting-rules}) and (\ref{counting-rules-y}). Meanwhile, the transverse derivatives that appear in other parts of computation (e.g. in the loops) and that are not accompanied by $y_T$, produce power suppressed terms.

In addition to the diagram A, one should take into account the diagrams that are of the same order of power counting. These are diagrams with radiation of collinear components of gluon, such as diagram B in fig.\ref{fig:diags}. Accounting of all such diagrams restores the QCD gauge invariance and equips the collinear and anti-collinear quark fields by half-infinite Wilson lines \cite{Belitsky:2002sm, Boer:2003cm}. Practically, it is convenient to impose the light-cone gauge for the background field. The gauge can be fixed for each collinear sector such that gluon fields with a $\sim \lambda^0$ power counting vanish, i.e. $A_{\bar n,+}=0$ and $A_{n,-}=0$. The detailed justification of such choice can be found e.g. in ref.\cite{Vladimirov:2021hdn}. For the present work, it is enough to know that this set of gauges provides the correct factorization.

Dropping the power-suppressed terms in eqn.(\ref{KPC:2}), one gets the hadronic tensor in the form
\begin{eqnarray}\label{KPC:3}
W^{\mu\nu}&=&\int \frac{d^4y}{(2\pi)^4}e^{-i(yq)}
\\\nn &&\quad \times 
\langle p_1|\langle p_2|\bar q_{\bar n}(y^-n+y_T) \gamma^\mu q_{n}(y^+\bar n+y_T) \bar q_{n}(0) \gamma^\nu q_{\bar n}(0) 
+\{\bar n\leftrightarrow n\}|p_1\rangle |p_2\rangle+...~.
\end{eqnarray}
The central assumption of the parton model is that a high-energy hadron consists of fields with corresponding collinear counting rules. Therefore, we can sort the fields with different collinearity into separate matrix elements. To do so, one needs to recouple the spinor and color indices, which could be done using the completeness relation of Dirac matrices
\begin{eqnarray}
A=\frac{1}{2}\sum_a \overline{\Gamma}_a A^{[\Gamma_a]},\qquad A^{[\Gamma]}=\frac{1}{2}\Tr\(A \Gamma\),
\end{eqnarray}
for any matrix $A$. Here, $\Gamma$ is a complete basis of Dirac matrices. The result reads
\begin{eqnarray}\nn
W^{\mu\nu}&=&
-\frac{1}{4N_c}\int \frac{d^4y}{(2\pi)^4}e^{-i(yq)}
\sum_{a,b}\Big(
\Tr\(\gamma^\mu \overline{\Gamma}_b\gamma^\nu \overline{\Gamma}_a\)
\widetilde{\Psi}_{\bar n}^{[\Gamma_a]}(y^-n+y_T)
\widetilde{\Psi}_{n}^{[\Gamma_b]}(-y^+\bar n-y_T)
\\&&\label{KPC:4}
\qquad\qquad
+
\Tr\(\gamma^\mu \overline{\Gamma}_a\gamma^\nu \overline{\Gamma}_b\)
\widetilde{\Psi}_{\bar n}^{[\Gamma_a]}(-y^-n-y_T)
\widetilde{\Psi}_{n}^{[\Gamma_b]}(y^+\bar n+y_T)\Big)+...,
\end{eqnarray}
where the common minus sing appears from the anti-commutation of the quark fields, and the factor $1/N_c$ is due to the recoupling of color indices. The matrix elements are
\begin{eqnarray}\label{def:Psi1}
\widetilde{\Psi}_{\bar n}^{[\Gamma]}(y^-n+y_T)=\langle p_1|\bar q_{\bar n}(y^-n+y_T)[y^-n+y_T,-\infty n+y_T]\frac{\Gamma}{2}[-\infty n,0]q_{\bar n}(0)|p_1\rangle,
\\\label{def:Psi1b}
\widetilde{\Psi}_{n}^{[\Gamma]}(y^+\bar n+y_T)=\langle p_2|\bar q_n(y^+\bar n+y_T)[y^+\bar n+y_T,-\infty \bar n+y_T]\frac{\Gamma}{2}[-\infty \bar n,0]q_n(0)|p_2\rangle.
\end{eqnarray}
These distributions are Fourier images of distributions $\widetilde{\Phi}$ (\ref{example:PHI}) over the variables $x$. In expressions (\ref{def:Psi1}, \ref{def:Psi1b}) the complete set of states in-between quark fields can be omitted because all distances are space-like.  

The expression (\ref{KPC:4}) is of indefinite power, because the TMD distribution $\Psi^{[\gamma^\mu]}$ is a mixture of terms with different power counting. It follows form eqn. (\ref{example:3}), where the first line is $\sim \lambda^0$, the second and fourth lines are $\sim \lambda^1$, etc. Therefore, the pure LP term is obtained by eliminating all, except the first, terms in eqn.(\ref{example:3}). It can be done by selecting only those $\Gamma$-matrices that project only ``good'' components. For the collinear part these matrices are
\begin{eqnarray}\label{def:Gamma+}
\Gamma^+_a=\{\gamma^+, \gamma^+\gamma^5, i\sigma^{\alpha+}\gamma^5\},
\end{eqnarray}
with $\alpha$ being transverse index. The corresponding elements of the basis are
\begin{eqnarray}\label{def:barGamma+}
\overline{\Gamma}^-_a=\{\gamma^-, -\gamma^-\gamma^5, -\frac{i}{2}\sigma^{\alpha-}\gamma^5\}.
\end{eqnarray}
For the decomposition in the anti-collinear sector one should use $\Gamma^-_a$ and $\overline{\Gamma}^+_a$ that are obtained by the replacement $n\leftrightarrow \bar n$ in (\ref{def:Gamma+}) and (\ref{def:barGamma+}), correspondingly. After restriction of $\Gamma$-matrices to the set (\ref{def:Gamma+}) the hadronic tensor contains only the LP contribution.

Finally, one passes to the momentum-fraction representation for TMD distributions and integrates over $y^\pm$. The result is
\begin{eqnarray}\label{KPC:5}
&&W^{\mu\nu}_{\text{LP}}=
-\frac{p_1^+p_2^-}{4N_c}\int \frac{d^2b}{(2\pi)^2}e^{-i(bq_T)}\int dx d\tilde x
\delta(xp_1^+-q^+)\delta(\tilde xp_2^--q^-)
\sum_{a,b}\Big(
\\&&\nn\quad
\Tr\(\gamma^\mu \overline{\Gamma}^+_b\gamma^\nu \overline{\Gamma}^-_a\)
\widetilde{\Phi}_{\bar n11}^{[\Gamma^+_a]}(x,b)
\widetilde{\Phi}_{n11}^{[\Gamma_b^-]}(-\tilde x,-b)
+
\Tr\(\gamma^\mu \overline{\Gamma}^-_a\gamma^\nu \overline{\Gamma}^+_b\)
\widetilde{\Phi}_{\bar n11}^{[\Gamma_a^+]}(-x,-b)
\widetilde{\Phi}_{n11}^{[\Gamma_b^-]}(\tilde x,b)\Big),
\end{eqnarray}
where $y_T$ is renamed as $b$, to match the standard notation. The expression (\ref{KPC:5}) is the celebrated expression for the LP TMD factorization. The TMD distributions with negative $x$ are related to the anti-quark distributions. 

This is a complete derivation of LP TMD factorization at the tree order. To go beyond LO, one should include interaction diagrams. Some examples are presented in fig.\ref{fig:diags}. The diagrams of type B (with a non-collinear gluon) or C do not contribute to the LP factorization. Indeed, any addition interaction with the hadrons (such as fig.\ref{fig:diags}B) increases the number of fields in the operator, and hence increases its dimension. The only exception is the radiation of collinear gluons, which is already accounted in the LP term. Therefore, the interactions that do not violate the power counting are possible only in-between the hard fields. Even so, some hard interactions are power suppressed, due to inhomogeneous counting rules for $y$. Indeed, the diagram \ref{fig:diags}C is $\sim 1/b^2$ in the position space, which leads to $\sim q_T^2/Q^2$ in the momentum space\footnote{
Since these power corrections appears due to the expansion of the propagator, they produces only even powers of $q_T/Q$-corrections. That describes the structure of eqn.(\ref{general-structure-1}).
}. Therefore, the perturbative correction at LP are given by the interactions in the vicinity of currents, such as the diagram \ref{fig:diags}D. As the result, the coefficient function is the product of coefficient functions for each current. It is in the one-to-one correspondence with the TMD-twist idea. Each semi-compact operator produces independent UV renormalization factor, that cancels the infrared poles of a corresponding part of coefficient function. This structure of pole cancellations is preserved at all powers of TMD factorization.

Finally, one should also take into account that separation of collinear and anti-collinear modes is ambiguous. For the small values of momenta the collinear and anti-collinear sectors overlap. There are several methods to avoid it. The double-counting of modes can be subtracted by means of the soft factor \cite{Manohar:2006nz, Collins:2011zzd} or rapidity renormalization factor \cite{Vladimirov:2017ksc}, or modes can be defined with explicit cut parameter that prevents the overlap \cite{Balitsky:2023hmh, Vladimirov:2021hdn}. The results coincide\footnote{
To reach a complete agreement, different schemes of the rapidity renomalization must  satisfy the same renormalization condition. Traditionally the TMDPDFs are defined such that the Drell-Yan cross-section does not incorporate any extra nonperturbative factor  \cite{Vladimirov:2017ksc}.
}. This procedure leads to an extra scales $\zeta$ and $\bar \zeta$, and corresponding evolution equations (\ref{evol:zeta}). 

\subsection{KPC for LP term at all orders}
\label{sec:KPC-all-order}

The extension of the TMD factorization beyond LP is conceptually straightforward (although very complicated technically). One should systematically expand the interaction vertices in the background fields, and sort operators over the twists and power counting. Detailed discussions are presented in refs.\cite{Vladimirov:2021hdn, Balitsky:2020jzt}. Performing this procedure, one receives the structure (\ref{general-structure-1}). In the following, I concentrate on the KPCs to the LP term. They are much simpler than other types of power corrections, because they incorporate  only the TMD distributions of twist-two. Let me consider the series of power corrections and extract only the twist-two component from it. This component produces the complete series of KPCs.

The key observation is that the TMD distributions of twist-two are given by only two-point operators. Indeed, an insertion of any extra field increases the dimension and, thus, the TMD-twist by at least one. The geometrical-twist-three operators are all three-point operators. Such a simple rule (twist $\sim$ number of fields) does not hold for higher-twist operators starting from twist-four, which could be three-point or four-point operators \cite{Braun:2009vc}. Still, one can formulate a rule: any two-point operator can contain geometrical twist-two part, but any three-point and higher-number-of-points operators necessarily have geometrical twist higher than two. It implies that any diagram with three or more particles in a single collinear sector (such as diagram B) does not contribute to KPCs for LP term. Consequently, the diagram A contains full information about twist-two part at all powers at LO in perturbative expansion.

Inspecting the LP derivation in sec.~\ref{sec:TMD-at-LP}, one finds that the power corrections were dropped in two places: the multipole decomposition (\ref{KPC:2}) and selection of ``good'' components of fields. If one does not perform these approximation the result will contain full information on KPCs. Lets release these approximations.

Starting with eqn.(\ref{KPC:1}) and performing the multipole expansion one obtains
\begin{eqnarray}\label{KPC:6}
&&W^{\mu\nu}=\int \frac{d^4y}{(2\pi)^4}e^{-i(yq)}
\sum_{n,m=0}^\infty \frac{(y^+)^n(y^-)^m}{n!m!}
\\ &&\nn\quad \times \langle p_1|\langle p_2|\bar q_{\bar n}(y^-n+y_T) \overleftarrow{\partial_-}^n \gamma^\mu \overrightarrow{\partial_+}^m q_{n}(y^+\bar n+y_T) J^\nu_{n\bar n}(0)
+\{\bar n\leftrightarrow n\}|p_1\rangle |p_2\rangle+...\,.
\end{eqnarray}
Here, the dots represent contributions with larger number of fields. The LP term (\ref{KPC:3}) is at $n=m=0$. Expressing the $\Gamma$-structure in the standard basis gives
\begin{eqnarray}\label{KPC:7}
&&W^{\mu\nu}=\frac{-1}{4N_c}\int \frac{d^4y}{(2\pi)^4}e^{-i(yq)}
\sum_{a,b}\sum_{n,m=0}^\infty \frac{(y^+)^n(y^-)^m}{n!m!}\Big[
\\\nn && \quad
\Tr\(\gamma^\mu \overline{\Gamma}_b\gamma^\nu \overline{\Gamma}_a\)~
\langle p_1|\overline{U}_q(y^-,y_T) \overleftarrow{\partial_-}^n \frac{\Gamma_a}{2} U_{q}(0,0_T)|p_1\rangle~
\langle p_2|\overline{U}_q(-y^+,-y_T)  \frac{\Gamma_b}{2} \overrightarrow{\partial_+}^m U_{q}(0,0_T)|p_2\rangle
\\\nn &&\qquad\qquad
+\{\bar n\leftrightarrow n\}\Big]+...\,,
\end{eqnarray}
where the I use the notation (\ref{example:U1}) to highlight the semi-compact operators. Note that operators in the matrix elements with $p_1$($p_2$) are (anti-)collinear. In contrast to the LP term, there are no constraints on $\Gamma$-matrices, since any Dirac component could produce a twist-two contribution, see example in eqn. (\ref{example:4}). 

The semi-compact operators in eqn.(\ref{KPC:7}) contain the mixture of TMD distribution, and should be decomposed over the definite twist contributions. For the pure $U_q$ operators the decomposition is given in eqn.(\ref{example:2}). However, the ``improper'' derivatives of $U_q$ (such as $\partial_-q_{\bar n}$) demolish this decomposition because they could not be presented as total derivatives of distributions. This issue is resolved by application of EOMs (\ref{def:EOMs}). The analysis of these operators is complicated in the general case, but essentially simpler if one seeks for only twist-one contribution. In this case, all gluon fields in EOMs could be dropped (since gluon-fields necessarily turn one-point operator to a higher-number-of-points operators), resulting to a simple rule
\begin{equation}\label{KPC:d-}
\partial_-\xi_{\bar n}=-\frac{\partial_T^2}{2\partial_+}\xi_{\bar n}+...,\qquad
\partial_-\eta_{\bar n}=\frac{\gamma^+\fnot \partial_T \partial_T^2}{4\partial^2_+}\xi_{\bar n}+...~,
\end{equation}
where dots represent the terms with two or larger number of fields. Here inverse derivatives are understood as the integral, alike (\ref{example-bad-comp}). Basically, the equation (\ref{KPC:d-}) tells that at the leading-twist approximation partons are  free massless fields. The higher powers of $\partial_-$ are
\begin{equation}\label{KPC:d-n}
\partial_-^n\xi_{\bar n}=\(-\frac{\partial_T^2}{2\partial_+}\)^n\xi_{\bar n}+...,\qquad
\partial^n_-\eta_{\bar n}=-\frac{\gamma^+\fnot \partial_T}{2\partial_+}\(-\frac{\partial_T^2}{2\partial_+}\)^n\xi_{\bar n}+...~.
\end{equation}
After these transformations, the fields $\xi$ could be promoted to the semi-compact operators $U_1$, by equipping them by semi-infinite Wilson lines, as it is demanded by the QCD gauge invariance. 

As a result of these manipulations, one obtains a long expression
\begin{eqnarray}\label{KPC:general0}
W_{\text{KPC}}^{\mu\nu}&=&
\frac{-1}{4N_c}\int \frac{d^4y}{(2\pi)^4}e^{-i(yq)}
\sum_{a,b}\sum_{n,m=0}^\infty \frac{(y^+)^{n}(y^-)^{m}}{n!m!}\frac{(-1)^{n+m}}{2^{n+m}}\Bigg\{
\\\nn 
&&
\Tr[\gamma_{T}^\mu\overline{\Gamma}_b^+ \gamma_{T}^{\nu}\overline{\Gamma}_a^-]\;
\frac{\partial_T^{2n}}{\partial_+^{n}}\widetilde{\Psi}_{\bar n11}^{[\Gamma_a^+]}\;
\frac{\partial_T^{2m}}{\partial_-^{m}}\widetilde{\Psi}_{n11}^{[\Gamma_b^-]}
\\\nn 
&&
-\(n^\nu \Tr[\gamma_{T}^\mu\overline{\Gamma}_b^+ \gamma_{T}^{\alpha}\overline{\Gamma}_a^-]
+n^\mu \Tr[\gamma_{T}^\alpha\overline{\Gamma}_b^+ \gamma_{T}^{\nu}\overline{\Gamma}_a^-]\)
\frac{\partial_\alpha}{\partial_+}
\frac{\partial_T^{2n}}{\partial_+^{n}}\widetilde{\Psi}_{\bar n11}^{[\Gamma_a^+]}\;
\frac{\partial_T^{2m}}{\partial_-^{m}}\widetilde{\Psi}_{n11}^{[\Gamma_b^-]}
\\\nn 
&&
-\(\bar n^\nu \Tr[\gamma_{T}^\mu\overline{\Gamma}_b^+ \gamma_{T}^{\alpha}\overline{\Gamma}_a^-]
+
\bar n^\mu \Tr[\gamma_{T}^\alpha\overline{\Gamma}_b^+ \gamma_{T}^{\nu}\overline{\Gamma}_a^-]\)
\frac{\partial_T^{2n}}{\partial_+^{n}}\widetilde{\Psi}_{\bar n11}^{[\Gamma_a^+]}\;
\frac{\partial_\alpha}{\partial_-}
\frac{\partial_T^{2m}}{\partial_-^{m}}\widetilde{\Psi}_{n11}^{[\Gamma_b^-]}
\\\nn
&&
-\frac{1}{2}\(
\Tr[\gamma_{T}^\mu\overline{\Gamma}_b^+ \gamma_{T}^{\beta}\gamma_T^\nu\gamma^\alpha_T \overline{\Gamma}_a^-]
+
\Tr[\gamma_{T}^{\alpha}\gamma_T^\mu\gamma^\beta_T\overline{\Gamma}_b^+\gamma_{T}^\nu  \overline{\Gamma}_a^-]\)
\frac{\partial_\alpha}{\partial_+}
\frac{\partial_T^{2n}}{\partial_+^{n}}\widetilde{\Psi}_{\bar n11}^{[\Gamma_a^+]}\;
\frac{\partial_\beta}{\partial_-}
\frac{\partial_T^{2m}}{\partial_-^{m}}\widetilde{\Psi}_{n11}^{[\Gamma_b^-]}
\\\nn 
&&
+\(
\bar n^\mu n^\nu \Tr[\gamma_{T}^\beta\overline{\Gamma}_b^+\gamma^\alpha_T\overline{\Gamma}_a^-]
+
n^\mu \bar n^\nu \Tr[\gamma_{T}^\alpha\overline{\Gamma}_b^+\gamma^\beta_T\overline{\Gamma}_a^-]\)
\frac{\partial_\alpha}{\partial_+}
\frac{\partial_T^{2n}}{\partial_+^{n}}\widetilde{\Psi}_{\bar n11}^{[\Gamma_a^+]}\;
\frac{\partial_\beta}{\partial_-}
\frac{\partial_T^{2m}}{\partial_-^{m}}\widetilde{\Psi}_{n11}^{[\Gamma_b^-]}
\\\nn 
&&
+
\Tr[\gamma_{T}^\alpha\overline{\Gamma}_b^+\gamma^\beta_T\overline{\Gamma}_a^-]
\(
\bar n^\mu \bar n^\nu 
\frac{\partial_T^{2n}}{\partial_+^{n}}\widetilde{\Psi}_{\bar n11}^{[\Gamma_a^+]}\;
\frac{\partial_\alpha}{\partial_-}
\frac{\partial_\beta}{\partial_-}
\frac{\partial_T^{2m}}{\partial_-^{m}}\widetilde{\Psi}_{n11}^{[\Gamma_b^-]}
+
n^\mu n^\nu 
\frac{\partial_\alpha}{\partial_+}
\frac{\partial_\beta}{\partial_+}
\frac{\partial_T^{2n}}{\partial_+^{n}}\widetilde{\Psi}_{\bar n11}^{[\Gamma_a^+]}\;
\frac{\partial_T^{2m}}{\partial_-^{m}}\widetilde{\Psi}_{n11}^{[\Gamma_b^-]}
\)
\\\nn 
&&
+\frac{1}{2}
\(
\bar n^\mu \Tr[\gamma_{T}^\sigma\overline{\Gamma}_b^+\gamma^\beta_T\gamma^\nu \gamma^\alpha\overline{\Gamma}_a^-]
+
\bar n^\nu \Tr[\gamma^\alpha_T\gamma_{T}^\mu\gamma^\beta_T\overline{\Gamma}_b^+\gamma^\sigma_T\overline{\Gamma}_a^-]\)
\frac{\partial_\alpha}{\partial_+}
\frac{\partial_T^{2n}}{\partial_+^{n}}\widetilde{\Psi}_{\bar n11}^{[\Gamma_a^+]}\;
\frac{\partial_\beta}{\partial_-}\frac{\partial_\sigma}{\partial_-}
\frac{\partial_T^{2m}}{\partial_-^{m}}\widetilde{\Psi}_{n11}^{[\Gamma_b^-]}
\\\nn 
&&
+\frac{1}{2}
\(
n^\mu \Tr[\gamma_{T}^\alpha\overline{\Gamma}_b^+\gamma^\sigma_T\gamma^\nu \gamma^\beta\overline{\Gamma}_a^-]
+
n^\nu \Tr[\gamma^\beta_T\gamma_{T}^\mu\gamma^\sigma_T\overline{\Gamma}_b^+\gamma^\alpha_T\overline{\Gamma}_a^-]\)
\frac{\partial_\alpha}{\partial_+}\frac{\partial_\beta}{\partial_+}
\frac{\partial_T^{2n}}{\partial_+^{n}}\widetilde{\Psi}_{\bar n11}^{[\Gamma_a^+]}\;
\frac{\partial_\sigma}{\partial_-}
\frac{\partial_T^{2m}}{\partial_-^{m}}\widetilde{\Psi}_{n11}^{[\Gamma_b^-]}
\\\nn
&&
+\frac{1}{4}
\Tr[\gamma_{T}^\alpha\gamma^\mu_T\gamma_T^\sigma \overline{\Gamma}_b^+\gamma^\rho_T\gamma^\nu \gamma_T^\beta\overline{\Gamma}_a^-]
\frac{\partial_\alpha}{\partial_+}\frac{\partial_\beta}{\partial_+}
\frac{\partial_T^{2n}}{\partial_+^{n}}\widetilde{\Psi}_{\bar n11}^{[\Gamma_a^+]}\;
\frac{\partial_\sigma}{\partial_-}\frac{\partial_\rho}{\partial_-}
\frac{\partial_T^{2m}}{\partial_-^{m}}\widetilde{\Psi}_{n11}^{[\Gamma_b^-]}\Bigg\}+(n\leftrightarrow \bar n),
\end{eqnarray}
where the argument of $\widetilde{\Psi}_{\bar n11}$ is $(y^-n+y_T)$, and the argument of $\widetilde{\Psi}_{n11}$ is $(-y^+\bar n-y_T)$. This expression contains all twist-two terms of TMD factorization that appear at LO in perturbation theory. The eliminated terms have higher twist.

Finally, one integrates over $y^\pm$ and obtains
\begin{eqnarray}\label{KPC:general1}
&&W_{\text{KPC}}^{\mu\nu}=
\frac{-p_1^+p_2^-}{4N_c}\int \frac{d^2b}{(2\pi)^2}e^{-i(bq_T)}\int dx d\tilde x \delta(xp_1^+-q^+)\delta(\tilde xp_2^--q^-)
\\\nn &&
\sum_{q,\bar q}\sum_{a,b}\sum_{n,m=0}^\infty \frac{\partial_{\tilde x}^{n}\partial_x^{m}}{n!m!}\frac{(-1)^{n+m}}{(2p_1^+p_2^-)^{n+m}}\frac{1}{x^n\tilde x^m}\Bigg\{
\\\nn 
&&
\Tr[\gamma_{T}^\mu\overline{\Gamma}_b^+ \gamma_{T}^{\nu}\overline{\Gamma}_a^-]\;
\partial_T^{2n}\widetilde{\Phi}_{\bar n11}^{[\Gamma_a^+]}\;
\partial_T^{2m}\widetilde{\Phi}_{n11}^{[\Gamma_b^-]}
\\\nn 
&&
+\frac{i}{xp_1^+}\(n^\nu \Tr[\gamma_{T}^\mu\overline{\Gamma}_b^+ \gamma_{T}^{\alpha}\overline{\Gamma}_a^-]
+n^\mu \Tr[\gamma_{T}^\alpha\overline{\Gamma}_b^+ \gamma_{T}^{\nu}\overline{\Gamma}_a^-]\)
\partial_\alpha \partial_T^{2n}
\widetilde{\Phi}_{\bar n11}^{[\Gamma_a^+]}\;
\partial_T^{2m}\widetilde{\Phi}_{n11}^{[\Gamma_b^-]}
\\\nn 
&&
+\frac{i}{\tilde xp_2^-}\(\bar n^\nu \Tr[\gamma_{T}^\mu\overline{\Gamma}_b^+ \gamma_{T}^{\alpha}\overline{\Gamma}_a^-]
+
\bar n^\mu \Tr[\gamma_{T}^\alpha\overline{\Gamma}_b^+ \gamma_{T}^{\nu}\overline{\Gamma}_a^-]\)
\partial_T^{2n}\widetilde{\Phi}_{\bar n11}^{[\Gamma_a^+]}\;
\partial_\alpha \partial_T^{2m}\widetilde{\Phi}_{n11}^{[\Gamma_b^-]}
\\\nn
&&
+\frac{1}{2x\tilde x p_1^+p_2^-}\Big(
\Tr[\gamma_{T}^\mu\overline{\Gamma}_b^+ \gamma_{T}^{\beta}\gamma_T^\nu\gamma^\alpha_T \overline{\Gamma}_a^-]
+
\Tr[\gamma_{T}^{\alpha}\gamma_T^\mu\gamma^\beta_T\overline{\Gamma}_b^+\gamma_{T}^\nu  \overline{\Gamma}_a^-]
\\\nn &&\qquad\qquad+
2\bar n^\mu n^\nu \Tr[\gamma_{T}^\beta\overline{\Gamma}_b^+\gamma^\alpha_T\overline{\Gamma}_a^-]
+
2n^\mu \bar n^\nu \Tr[\gamma_{T}^\alpha\overline{\Gamma}_b^+\gamma^\beta_T\overline{\Gamma}_a^-]\Big)
\partial_\alpha
\partial_T^{2n}\widetilde{\Phi}_{\bar n11}^{[\Gamma_a^+]}\;
\partial_\beta
\partial_T^{2m}\widetilde{\Phi}_{n11}^{[\Gamma_b^-]}
\\\nn 
&&
-
\Tr[\gamma_{T}^\alpha\overline{\Gamma}_b^+\gamma^\beta_T\overline{\Gamma}_a^-]
\(
\frac{\bar n^\mu \bar n^\nu }{\tilde x^2 (p_2^-)^2}
\partial_T^{2n}\widetilde{\Phi}_{\bar n11}^{[\Gamma_a^+]}\;
\partial_\alpha
\partial_\beta
\partial_T^{2m}\widetilde{\Phi}_{n11}^{[\Gamma_b^-]}
+
\frac{n^\mu n^\nu }{x^2 (p_1^+)^2}
\partial_\alpha
\partial_\beta
\partial_T^{2n}\widetilde{\Phi}_{\bar n11}^{[\Gamma_a^+]}\;
\partial_T^{2m}\widetilde{\Phi}_{n11}^{[\Gamma_b^-]}
\)
\\\nn 
&&
+\frac{i}{2 x\tilde x^2 p_1^+(p_2^-)^2}
\(
\bar n^\mu \Tr[\gamma_{T}^\sigma\overline{\Gamma}_b^+\gamma^\beta_T\gamma^\nu \gamma^\alpha\overline{\Gamma}_a^-]
+
\bar n^\nu \Tr[\gamma^\alpha_T\gamma_{T}^\mu\gamma^\beta_T\overline{\Gamma}_b^+\gamma^\sigma_T\overline{\Gamma}_a^-]\)
\partial_\alpha
\partial_T^{2n}\widetilde{\Phi}_{\bar n11}^{[\Gamma_a^+]}\;
\partial_\beta\partial_\sigma
\partial_T^{2m}\widetilde{\Phi}_{n11}^{[\Gamma_b^-]}
\\\nn 
&&
+\frac{i}{2 x^2\tilde x (p_1^+)^2p_2^-}
\(
n^\mu \Tr[\gamma_{T}^\alpha\overline{\Gamma}_b^+\gamma^\sigma_T\gamma^\nu \gamma^\beta\overline{\Gamma}_a^-]
+
n^\nu \Tr[\gamma^\beta_T\gamma_{T}^\mu\gamma^\sigma_T\overline{\Gamma}_b^+\gamma^\alpha_T\overline{\Gamma}_a^-]\)
\partial_\alpha\partial_\beta
\partial_T^{2n}\widetilde{\Phi}_{\bar n11}^{[\Gamma_a^+]}\;
\partial_\sigma
\partial_T^{2m}\widetilde{\Phi}_{n11}^{[\Gamma_b^-]}
\\\nn
&&
+\frac{1}{4(x\tilde x p_1^+p_2^-)^2}
\Tr[\gamma_{T}^\alpha\gamma^\mu_T\gamma_T^\sigma \overline{\Gamma}_b^+\gamma^\rho_T\gamma^\nu \gamma_T^\beta\overline{\Gamma}_a^-]
\partial_\alpha\partial_\beta
\partial_T^{2n}\widetilde{\Phi}_{\bar n11}^{[\Gamma_a^+]}\;
\partial_\sigma \partial_{\rho}
\partial_T^{2m}\widetilde{\Phi}_{n11}^{[\Gamma_b^-]}\Bigg\},
\end{eqnarray}
where arguments of the TMD-distributions are $(x,b)$ for $\widetilde{\Phi}_{\bar n11}$ and $(\tilde x,b)$ for $\widetilde{\Phi}_{n11}$. The derivatives $\partial_x=\frac{\partial}{\partial x}$ ($\partial_{\tilde x}=\frac{\partial}{\partial\tilde x}$) comes from from the integration over $y^-$($y^+$). Herewith, distributions $\widetilde{\Phi}_{\bar n11}$ are quark distributions and $\widetilde{\Phi}_{n11}$ are anti-quark distributions. The sum $\sum_{q,\bar q}$ indicates the addition of the term with quark and anti-quark distributions exchanged. This expression is valid for any polarizations. The example of unpolarized cases are given in sec.~\ref{sec:structure-general-main}. In particular, the LP expression is given in eqn.(\ref{LP:unpol-example-position}), NLP in eqn.(\ref{NLP:unpol-example-position}), and NNLP in eqn.(\ref{NNLP:unpol-example-position}). The NLP part of eqn.(\ref{KPC:general1}) exactly reproduces the KPC-part of the full NLP expression derived in ref.\cite{Vladimirov:2021hdn, Balitsky:2020jzt}.

It is complicated to investigate the properties (\ref{KPC:general1}) in the general form. However, one can confirm that the expression (\ref{KPC:general1}) is exactly EM-gauge invariant,
\begin{eqnarray}\label{qW=0}
q_\mu W_{\text{KPC}}^{\mu\nu}=0.
\end{eqnarray}
The truncated at N$^n$LP order part of series (\ref{KPC:general1}) is not transverse, but violates EM-gauge invariance up to order N$^{n+1}$LP. The exact restoration involves all terms. Moreover, the series (\ref{KPC:general1}) could not be split into independently gauge-invariant parts or at least such a split is not natural. The terms with derivatives $\partial_x$ and $\partial_{\tilde x}$ play important role in the restoration of gauge invariance, since they (after integration by parts) turn $q^\pm$ to $p^\pm$, which is required for the cancellation between different orders. The expression (\ref{KPC:general1}) is also frame-invariant, although it is complicated to confirm. These invariances are discussed explicitly with the unpolarized example in sec.~\ref{sec:EM-example} and sec.~\ref{sec:frame-example}, and they are obvious in the summed form (\ref{KPC-main}).

\subsection{KPCs beyond LO and the argument of the coefficient function}
\label{sec:argument-Q}

In the present approximation (solely twist-two TMD operators, neglecting $q_T/Q$ corrections) the perturbative corrections come only from the corrections to the EM currents. Indeed, an inclusion of a non-collinear external gluon increases the twist of operators, and the in-between currents interaction (alike fig.\ref{fig:diags}C) induces $q_T/Q$ corrections. Therefore, one can consider perturbative corrections for the currents $J_{\bar n n}$ and $J_{n \bar n}$ (\ref{KPC:J}) independently. The NLO LP computation the coefficient function is performed in multiple works, see f.i. refs.\cite{Collins:2011zzd, Echevarria:2011epo, Becher:2010tm, Manohar:2003vb, Vladimirov:2021hdn}. Presently, it is known up to N$^4$LO accuracy (4-loop) \cite{Lee:2022nhh, Moult:2022xzt}. The NLP NLO computation has been done explicitly in ref.\cite{Vladimirov:2021hdn}, and the result coincides with LP. It is a non-trivial check since, intermediate expressions and the algebra of NLP computation are different from those at LP (see the detailed discussion in sec.6 of ref.\cite{Vladimirov:2021hdn}). As it is explained above, the coefficient function of KPC must be the same as the LP coefficient function. In this section I would like to confirm it by explicit computation.

The perturbative computations beyond LP are much simpler in the position space. It is because Feynman propagators in the position space are free of external momenta, which appear only in the numerator of diagrams. Therefore, one should not worry about increasing singularity of power suppressed parts of integrals as it happen in the momentum space. The explanation of convenient technique for loop-computation with multiple examples can be found f.i. in refs.~\cite{Vladimirov:2021hdn, Balitsky:1987bk, Scimemi:2019gge, Braun:2021aon}. 

The NLO correction for EM current is given by the diagram shown in fig.\ref{fig:diags} D. In the position space, this diagram (here for $J_{\bar n n}(0)$) reads 
\begin{eqnarray}
\text{diag}&=&g^2 C_F\frac{\Gamma^2(2-\epsilon)\Gamma(1-\epsilon)}{16\pi^{d/2}}\int d^dxd^dz \frac{\bar q_{\bar n}(x)\gamma^\nu\fnot x \gamma^\mu \fnot z \gamma_\nu q_n(z)}{[-x^2+i0]^{2-\epsilon}[-z^2+i0]^{2-\epsilon}[-(x-z)^2+i0]^{1-\epsilon}},~~
\end{eqnarray}
where $C_F=(N_c^2-1)/(2N_c)$ and $d=4-2\epsilon$ is the parameter of the dimensional regularization. This expression incorporates operators of all powers and twists. The power expansion is obtained from the decomposition of external quark fields over ``good'' and ``bad'' components, and their expansion in the vicinity of the collinear direction. Importantly, the power expansion can be done at the level of the integrand (since integral is convergent due to the dimensional regularization). In this way, the only complications of higher-power computation is the increasing (from power to power) size of the numerator.

The LP and NLP computation has been carried out in ref.\cite{Vladimirov:2021hdn}. It is convenient to write the result in the mixed momentum-coordinate representation
\begin{eqnarray}\label{KPC:coeff-LP}
\text{diag}_{\text{LP}}&=&\frac{a_s C(\epsilon)}{(-2k_1^+k_2^-)^\epsilon}\bar \xi_{\bar n}\gamma_T^\mu \xi_n,
\\\label{KPC:coeff-NLP}
\text{diag}_{\text{NLP}}&=&\frac{a_s C(\epsilon)}{(-2k_1^+k_2^-)^\epsilon}\(-n^\mu\bar \xi_{\bar n}\frac{\overleftarrow{\fnot \partial}_T}{\overleftarrow{\partial}_+} \xi_n
-\bar n^\mu\bar \xi_{\bar n}\frac{\overrightarrow{\fnot \partial}_T}{\overrightarrow{\partial}_+} \xi_n\)
\end{eqnarray}
where 
\begin{eqnarray}
C(\epsilon)=2C_F\frac{\Gamma(\epsilon)\Gamma(-\epsilon)\Gamma(2-\epsilon)}{\Gamma(3-2\epsilon)}(2-\epsilon+2\epsilon^2),
\end{eqnarray}
and $k_1$ and $k_2$ are the momenta of quark fields $\bar q_{\bar n}$ and $q_n$ (in the position space they appear as derivatives). The terms (\ref{KPC:coeff-LP}) and (\ref{KPC:coeff-NLP}) provide the coefficient function for the LP and NLP parts EM-current. In this way, one confirms that the coefficient function of KPC at NLP is the same as for LP term (at least at NLO). 

The coefficient $C(\epsilon)$ contains infra-red divergences (since $J^\mu$ is a conserved current, $C(\epsilon)$ is UV finite). These divergences are compensated by the UV renormalization factors of operators $U_1$. These factors are $Z_{U1}$ for $U_1$ (anti-collinear) and $Z_{U1}^\dagger$ for $\overline U_1$ (collinear). These factors contain collinear divergence, which is removed by the analogous divergence in the rapidity renormalization. The LO expression for $Z_{U1}$ is 
\begin{eqnarray}
Z_{U1}(\zeta)=1+\frac{a_sC_F}{\epsilon}\(\frac{1}{\epsilon}+\frac{3}{2}+\ln\(\frac{\mu^2}{\zeta}\)\)+\mathcal{O}(a_s^2),
\end{eqnarray}
where $\zeta$ is the scale of the rapidity renormalization. Performing the renormalization procedure, one obtains
\begin{eqnarray}\label{C1ZZ=CV}
(1+a_s C(\epsilon)+\mathcal{O}(a_s^2))Z^\dagger_{U1}(\zeta)Z_{U1}(\bar \zeta)=C_V,
\end{eqnarray} 
where $a_s=g^2/(4\pi)^2$. Importantly, for the cancellation of poles one must impose $\zeta\bar \zeta=(2q^+q^-)^2$, where it is used that $k_1^+=q^+$ and $k_2^-=q^-$ due to the $\delta$-functions (\ref{KPC:5}). The coefficient function $C_V$ is
\begin{eqnarray}\label{coeff-NLO}
C_V=1+a_s C_F\(-\mathbf{L}^2+3 \mathbf{L}-8+\frac{\pi^2}{6}\)+\mathcal{O}(a_s^2),
\end{eqnarray}
where $\mathbf{L}=\ln(-2q^+q^-/\mu^2)$. The coefficient function is multiplicative. The product of current has the coefficient function 
\begin{eqnarray}
C_0=|C_V|^2=1+2a_s C_F\(-|\mathbf{L}|^2+3 |\mathbf{L}|-8+\frac{7\pi^2}{6}\)+\mathcal{O}(a_s^2),
\end{eqnarray}
where it has been taken into account that $q^+q^->0$. 

Let me emphasize that the argument of the coefficient function is $2q^+q^-$. This is the result of the formal computation in the factorization approach (see e.g. \cite{Vladimirov:2021hdn, Manohar:2003vb}). Often, the argument of coefficient function at LP is replaced by $q^2$ for simplicity. However, it is not entirely correct since $q^2=2q^+q^-+q_T^2$ has an indefinite power counting.

The NLO computation can be easily automatized (necessary integrals in general form are given in appendix B or ref.\cite{Vladimirov:2021hdn}). Using the package \textit{FeynCalc} \cite{Shtabovenko:2020gxv}, I have generated the first few powers\footnote{
I thank Oscar del Rio for the cross-check of the N$^2$LP computation.
}. They are
\begin{eqnarray}\label{KPC:coeff-series}
\text{diag}_{\text{N$^2$LP}}&=&\frac{a_s C(\epsilon)}{(-2k_1^+k_2^-)^\epsilon}\Big[
-\frac{1}{2}
\bar \xi_{\bar n}\frac{\overleftarrow{\fnot \partial}_T}{\overleftarrow{\partial}_+}\gamma_T^\mu \frac{\overrightarrow{\fnot \partial}_T}{\overrightarrow{\partial}_+}\xi_n
-\epsilon \frac{(k_{1}k_2)_T}{k_1^+k_2^-}\bar \xi_{\bar n}\gamma_T^\mu \xi_n
\Big],
\\\nn
\text{diag}_{\text{N$^3$LP}}&=&\frac{a_s C(\epsilon)}{(-2k_1^+k_2^-)^\epsilon}\(-\epsilon \frac{(k_{1}k_2)_T}{k_1^+k_2^-}\)\(-n^\mu\bar \xi_{\bar n}\frac{\overleftarrow{\fnot \partial}_T}{\overleftarrow{\partial}_+} \xi_n
-\bar n^\mu\bar \xi_{\bar n}\frac{\overrightarrow{\fnot \partial}_T}{\overrightarrow{\partial}_+} \xi_n\),
\\\nn
\text{diag}_{\text{N$^4$LP}}&=&\frac{a_s C(\epsilon)}{(-2k_1^+k_2^-)^\epsilon}\Big[
\frac{\epsilon}{2} \frac{(k_{1}k_2)_T}{k_1^+k_2^-}
\bar \xi_{\bar n}\frac{\overleftarrow{\fnot \partial}_T}{\overleftarrow{\partial}_+}\gamma_T^\mu \frac{\overrightarrow{\fnot \partial}_T}{\overrightarrow{\partial}_+}\xi_n
+\(\frac{\epsilon(1+\epsilon)}{2} \frac{(k_{1}k_2)_T}{k_1^+k_2^-}-
\frac{\epsilon}{4} \frac{k_{1T}^2k_{2T}^2}{(k_1^+k_2^-)^2}\)\bar \xi_{\bar n}\gamma_T^\mu \xi_n
\Big],
\\\nn
\text{diag}_{\text{N$^5$LP}}&=&\frac{a_s C(\epsilon)}{(-2k_1^+k_2^-)^\epsilon}\(\frac{\epsilon(1+\epsilon)}{2} \frac{(k_{1}k_2)^2_T}{(k_1^+k_2^-)^2}-
\frac{\epsilon}{4} \frac{k_{1T}^2k_{2T}^2}{(k_1^+k_2^-)^2}\)\(-n^\mu\bar \xi_{\bar n}\frac{\overleftarrow{\fnot \partial}_T}{\overleftarrow{\partial}_+} \xi_n
-\bar n^\mu\bar \xi_{\bar n}\frac{\overrightarrow{\fnot \partial}_T}{\overrightarrow{\partial}_+} \xi_n\)
,
\\\nn
\text{diag}_{\text{N$^6$LP}}&=&
\frac{a_s C(\epsilon)}{(-2k_1^+k_2^-)^\epsilon}\Big[
\(\frac{\epsilon(1+\epsilon)}{2} \frac{(k_{1}k_2)^2_T}{(k_1^+k_2^-)^2}-
\frac{\epsilon}{4} \frac{k_{1T}^2k_{2T}^2}{(k_1^+k_2^-)^2}\)
\bar \xi_{\bar n}\frac{\overleftarrow{\fnot \partial}_T}{\overleftarrow{\partial}_+}\gamma_T^\mu \frac{\overrightarrow{\fnot \partial}_T}{\overrightarrow{\partial}_+}\xi_n
\\\nn &&
+\(-\frac{\epsilon(1+\epsilon)(2+\epsilon)}{6} \frac{(k_{1}k_2)^3_T}{(k_1^+k_2^-)^3}-
\frac{\epsilon(1+\epsilon)}{4} \frac{k_{1T}^2k_{2T}^2(k_1k_2)_T}{(k_1^+k_2^-)^3}\)\bar \xi_{\bar n}\gamma_T^\mu \xi_n
\Big],
\end{eqnarray}
and so on. For the convincingness, the computation has been performed up to N$^{10}$LP, and it reproduces the presented pattern. 

At the first glance, expressions (\ref{KPC:coeff-series}) demonstrate the violation of the factorization theorem. Indeed, naively, one should expect to obtain the effective current (\ref{KPC:J-decompose}) multiplied by $C(\epsilon)$. I.e. all terms except the first for $\text{diag}_{\text{N$^2$LP}}$ should vanish. Instead, the expressions contain the new terms, equipped by different coefficients. It is especially confusing that these new terms have different order of $\epsilon$ singularities. Due to the latter, these terms cannot be renormalized by $Z_{U1}$'s (\ref{C1ZZ=CV}), and the factorization theorem seems to be violated at NNLP and higher.

Further inspection of corrections (\ref{KPC:coeff-series}) reveals a certain pattern. The all-power series (\ref{KPC:coeff-series}) can be rewritten as
\begin{eqnarray}\label{coeff:argument}
\sum_{k=0}^\infty \text{diag}_{\text{N$^k$LP}}&=&
\frac{a_s C(\epsilon)}{\(-2k_1^+k_2^--2(k_1k_2)_T-\frac{k_{1T}^2k_{2T}^2}{2k_1^+k_2^-}\)^\epsilon} J^{\text{tw2}}_{\bar n n},
\end{eqnarray}
where 
\begin{eqnarray}\label{KPC:J-decompose}
J^{\text{tw2}}_{\bar nn}&=&\bar \xi_{\bar n}\gamma_T^\mu \xi_n
-n^\mu\bar \xi_{\bar n}\frac{\overleftarrow{\fnot \partial}_T}{\overleftarrow{\partial}_+} \xi_n
-\bar n^\mu\bar \xi_{\bar n}\frac{\overrightarrow{\fnot \partial}_T}{\overrightarrow{\partial}_+} \xi_n
-\frac{1}{2}
\bar \xi_{\bar n}\frac{\overleftarrow{\fnot \partial}_T}{\overleftarrow{\partial}_+}\gamma_T^\mu \frac{\overrightarrow{\fnot \partial}_T}{\overrightarrow{\partial}_+}\xi_n,
\end{eqnarray}
is the twist-two part of the EM current. Therefore, the coefficient function for all KPCs \textit{is the same, but has a different argument}. The argument of the coefficient function is 
\begin{eqnarray}\label{def:X}
X=(2q^+q^--2(k_1k_2)_T-k_{1T}^2k_{2T}^2/(2q^+q^-)),
\end{eqnarray}
instead of $2q^+q^-$, that appears at LP and NLP cases. The cancellation of poles also imposes 
$$\zeta\bar \zeta=X^2.$$ The coefficient function $C_0(X)$ can be used only as the whole, because any strict fixed-power expansion would lead to the non-cancellation of poles between infra-red poles of coefficient function, and UV poles of twist-one operators. To my best knowledge, it is the first example of accumulation of power corrections into the argument of the coefficient function in factorization theorems. In sec.~\ref{sec:W-summed}, it is shown that $X^2$ turns to $Q^2$ once KPCs are summed.

In the position space the variables $k_{1,2}$ are represented by the differential operators. Therefore, in the position space, the coefficient function (\ref{coeff:argument}) is a differential operator in the transverse space. That is the main reasons to switch to the momentum space in the following discussions.

Another contribution that reveals only beyond LO is the restoration of the boost-invariance. The boost-invariance demands that the factorization theorem is independent of the values of $\zeta$'s as far as $\zeta \bar \zeta=X^2$ (this can be demonstrated using different approaches \cite{Collins:2012ss, Echevarria:2012js, Vladimirov:2017ksc}). Therefore, the factorized expression should be invariant under
\begin{eqnarray}\label{boost-invariance}
\zeta \to \alpha \zeta,\qquad \bar \zeta\to \alpha^{-1}\bar \zeta,
\end{eqnarray}
at each given power. The expression (\ref{KPC:general1}) does not preserve this property. Indeed, the combinations with transverse derivatives, such as $\Phi_{\bar n}(\zeta)\partial_{T\mu} \Phi_{n}(\bar \zeta)$ are not invariant but produces the terms $\sim (\alpha-1)\partial_\mu\mathcal{D}$ (for $\alpha\sim 1$), where $\mathcal{D}$ is the Collins-Soper kernel (\ref{evol:zeta}). 

The restoration of the boost-invariance happens with the accounting of the higher-twist terms. The higher-twist distributions appears in the factorized expressions with the divergent convolution. These divergences are called the special rapidity divergences \cite{Rodini:2022wki}. They are somewhat similar to the end-point divergences in the heavy-quark factorizations \cite{Beneke:2001at, Beneke:2003zv}, but have a different nature, and are rapidity divergences according to the general criterion \cite{Vladimirov:2017ksc}. The leading contribution of these divergences is proportional to the derivative of the Collins-Soper kernel and twist-two TMD distributions. In the sum of all terms these divergences cancel, and leave the remnant proportional to $\ln(\zeta/\bar \zeta)$, which restores the boost-invariance. Practically, it leads to the replacement
\begin{eqnarray}\label{d->D}
\partial_{T\mu}\Phi_{n}\to D_\mu\(\frac{\bar \zeta}{\zeta}\)\Phi_{n}=\[\frac{\partial}{\partial b^\mu}-\frac{1}{2}\(\frac{\partial \mathcal{D}(b,\mu)}{\partial b^\mu}\)\ln\(\frac{\bar \zeta}{\zeta}\)\]\Phi_{n},
\end{eqnarray}
and analogous for $\Phi_{\bar n}$. Here,
\begin{eqnarray}\label{def:D}
D_\rho\(\frac{\bar \zeta}{\zeta}\)=\frac{\partial}{\partial b^\rho}-\frac{1}{2}\(\frac{\partial \mathcal{D}(b,\mu)}{\partial b^\rho}\)\ln\(\frac{\bar \zeta}{\zeta}\).
\end{eqnarray}
It is straightforward to check that the structure $\Phi_{\bar n}(\zeta)D_{\mu}(\bar \zeta/\zeta) \Phi_{n}(\bar \zeta)$ is invariant under (\ref{boost-invariance})

The mechanism of special rapidity divergences is well-understood at NLP, \cite{Rodini:2022wic, Rodini:2023plb}, but not beyond. The explicit check requires the computation with increased number of loops for each next power. The terms $\sim (\partial_T \mathcal{D})^n$, which required at $n$'th order, are $\mathcal{O}(a_s^n)$ in perturbative order and could be obtained only with $n$-loop computation. Also, one needs the complete higher-twist structure to extract the divergences. Such computation goes far beyond the present work\footnote{
I have checked explicitly that the cancellation of special-rapidity divergences between twist-three part of the NNLP terms, which are indicated as $D\Phi_{\mathbf{2}}\times \Phi_{\mathbf{3}}$ and $\Phi_{\mathbf{3}}\times \Phi_{\mathbf{3}}$ in the expression (\ref{general-structure-1}). These terms are relatively simple to obtain using NLP factorization. I have found that their one-loop special rapidity divergences produce the terms required for restoration of boost-invariance. The performed check in incomplete (and for that reason is not presented here explicitly), because it was done using only one-loop computation and caught only the terms $\sim \partial_T \mathcal{D}$, but does not catch terms $\sim (\partial_T \mathcal{D})^2$.
}. Nonetheless, it seems safe to extend the NLP case to all powers and declare that the complete expression for KPCs (\ref{KPC:general1}) requires the replacement (\ref{d->D}). This replacement restores the boost-invariance (\ref{boost-invariance}) at each order of power expansion independently.

In this section, I have demonstrated at NLO that the coefficient functions for all KPCs are the same. It is in the full agreement with the expectations, since KPCs must reproduce LP term, in order to preserve EM-gauge invariance (discussed in sec.~\ref{sec:EM-example}). However, it appears that the coefficient function has the mixed-power argument (as shown in eqn.(\ref{coeff:argument})). Also, I argue that the transverse derivatives must be replaced by the boost-invariant transverse derivatives (\ref{d->D}). Equipped by these modifications the expression (\ref{KPC:general1}) represents the all-power part of the TMD factorization that contains solely twist-two distributions, without $q_T/Q$ corrections.

\section{Summation of KPC series}
\label{sec:sumKPC}

The series of KPCs can be summed to a simpler expression. The direct summation of expression (\ref{KPC:coeff-series}) is possible but very tedious. The main complication is the derivatives over $x$ and $\tilde x$ which produce generating functions with complicated argument (see (\ref{def:xi})). The simpler way is to sum the multipole expansion separately for each collinear sector before the tensor decomposition. It leads a simple and intuitive picture of KPCs as the scattering of free massless partons. The details of computation are presented in this section.

\subsection{Twist-two part of a general TMD correlator}
\label{sec:tw2-general}

The procedure of the assembling fields into matrix elements and the multipole expansion are commuting operations. Due to it, one can collect the fields of different modes into matrix elements already in eqn.(\ref{KPC:1}). The result is (compare with (\ref{KPC:4}))
\begin{eqnarray}\nn
W^{\mu\nu}&=&
-\frac{1}{4N_c}\int \frac{d^4y}{(2\pi)^4}e^{-i(yq)}
\sum_{a,b}\Big(
\Tr\(\gamma^\mu \overline{\Gamma}_b\gamma^\nu \overline{\Gamma}_a\)
\widetilde{\Psi}_{\bar n}^{[\Gamma_a]}(y)
\widetilde{\Psi}_{n}^{[\Gamma_b]}(-y)
\\&&\label{sumKPC:1}
\qquad\qquad
+
\Tr\(\gamma^\mu \overline{\Gamma}_a\gamma^\nu \overline{\Gamma}_b\)
\widetilde{\Psi}_{\bar n}^{[\Gamma_a]}(-y)
\widetilde{\Psi}_{n}^{[\Gamma_b]}(y)\Big),
\end{eqnarray}
where $\Psi(y)$ are the generalization of (\ref{def:Psi1}) for a four-dimensional argument,
\begin{eqnarray}\label{def:Psi2}
\widetilde{\Psi}_{\bar n}^{[\Gamma]}(y)=\langle p_1|\bar q_{\bar n}(y)[y,-\infty n+y_T+\bar ny^+]\frac{\Gamma}{2}[-\infty n,0]q_{\bar n}(0)|p_1\rangle,
\\
\widetilde{\Psi}_{n}^{[\Gamma]}(y)=\langle p_2|\bar q_n(y)[y,-\infty \bar n+y_T+ny^-]\frac{\Gamma}{2}[-\infty \bar n,0]q_n(0)|p_2\rangle.
\end{eqnarray}
These matrix elements are alike ordinary TMD distributions, but with a generally-valued separation between semi-compact operators. The direction of Wilson lines is preserved.

The expression (\ref{def:Psi2}) is not a factorization theorem, but only an intermediate expression that is used to sum KPCs. The functions $\Psi$ should be understood as the generating functions for the series of power corrections (\ref{KPC:7}). I.e
\begin{eqnarray}\label{sumKPC:2}
\widetilde{\Psi}_{\bar n}^{[\Gamma]}(y)=\sum_{n=0}^\infty \frac{(y^+)^n}{n!}[\partial_-^n\widetilde{\Psi}_{\bar n}^{[\Gamma]}(y^- n+y_T)],
\end{eqnarray}
where
\begin{eqnarray}
[\partial_-^n\widetilde{\Psi}_{\bar n}^{[\Gamma]}(y^- n+y_T)]=
\langle p_1|\bar q_{\bar n}(y^- n+y_T)[y^- n+y_T,-\infty n+y_T]\overleftarrow{\partial}_-^n\frac{\Gamma}{2}[-\infty n,0]q_{\bar n}(0)|p_1\rangle,
\end{eqnarray}
and similar for $\Psi_{n}$. Substituting this expression into eqn.(\ref{sumKPC:1}) and making the twist-decomposition one gets eqn.(\ref{KPC:7}). 

The function $\Psi$ is an infinite series of TMD distributions of different twists. The twist-two part of (\ref{sumKPC:2}) can be extracted term-by-term using EOMs (\ref{def:EOMs}, \ref{KPC:d-}) in the same way as it was done in sec.~\ref{sec:KPC-all-order}. It is convenient to present the result in the following form
\begin{eqnarray}\label{sumKPC:3}
\Big[\widetilde{\Psi}_{\bar n}^{[\Gamma]}(y)\Big]_{\text{tw-2}}&=&
\sum_{n=0}^\infty \frac{(y^+)^n}{n!}\(\frac{-\partial_T^2}{2\partial_+}\)^n\Big[
\widetilde{\Psi}_{\bar n11}^{\llbracket \Gamma \rrbracket}(y^-n+y_T)
-\frac{\partial_{T\alpha }}{2\partial_+}
\widetilde{\Psi}_{\bar n11}^{\llbracket \gamma^\alpha\gamma^+\Gamma+\Gamma\gamma^+\gamma^\alpha \rrbracket}(y^-n+y_T)
\\\nn &&
+\frac{\partial_{T\alpha}\partial_{T\beta}}{4\partial_+^2}
\widetilde{\Psi}_{\bar n11}^{\llbracket \gamma^\alpha\gamma^+\Gamma\gamma^+\gamma^\beta \rrbracket}(y^-n+y_T)\Big],
\end{eqnarray}
where $\partial_{T\alpha}=\frac{\partial}{\partial y_T^\alpha}$, $\partial_+=\frac{\partial}{\partial y_-}$ and
\begin{eqnarray}
\llbracket \Gamma \rrbracket=[\frac{\gamma^+\gamma^-}{2}\Gamma\frac{\gamma^-\gamma^+}{2}].
\end{eqnarray}
The dropped terms are the distributions of twist-three and higher. Note that on the level of the cross-section the derivatives turns to the boost-invariant derivatives (\ref{def:D}), and thus the expression satisfy the leading-twist TMD evolution equation (\ref{evol:mu}, \ref{evol:zeta}).

The summation of series (\ref{sumKPC:3}) is straightforward in the momentum space, where it reads
\begin{eqnarray}\label{sumKPC:4}
\Big[\widetilde{\Psi}_{\bar n}^{[\Gamma]}(y)\Big]_{\text{tw-2}}&=&
\sum_{n=0}^\infty \frac{p^+}{n!}
\int dx d^2 k_T e^{ixp^+ y^-+i(ky)_T}\(\frac{-iy^+ k_T^2}{2xp^+}\)^n\Big[
\\\nn &&
\Phi_{\bar n11}^{\llbracket \Gamma \rrbracket}(x,k_T)
-\frac{k_{T\alpha }}{2xp^+}
\Phi_{\bar n11}^{\llbracket \gamma^\alpha\gamma^+\Gamma+\Gamma\gamma^+\gamma^\alpha \rrbracket}(x,k_T)
+\frac{k_{T\alpha}k_{T\beta}}{4(xp^+)^2}
\Phi_{\bar n11}^{\llbracket \gamma^\alpha\gamma^+\Gamma\gamma^+\gamma^\beta \rrbracket}(x,k_T)\Big].
\end{eqnarray}
The sum over $n$ produces the exponent $\exp(-i y^+ k_T^2/(2xp^+))$. Introducing an auxiliary variable $k^-=-k_T^2/(2xp^+)$, one can rewrite (\ref{sumKPC:4}) in a compact form 
\begin{eqnarray}\label{sumKPC:5}
\Big[\widetilde{\Psi}_{\bar n}^{[\Gamma]}(y)\Big]_{\text{tw-2}}&=&
p^+
\int dx d^4 k \,(2k^+)e^{i(ky)}\delta(k^+-xp^+)\delta(k^2)\Big[
\\\nn &&
\Phi_{\bar n11}^{\llbracket \Gamma \rrbracket}(x,k_T)
-\frac{k_{T\alpha }}{2xp^+}
\Phi_{\bar n11}^{\llbracket \gamma^\alpha\gamma^+\Gamma+\Gamma\gamma^+\gamma^\alpha \rrbracket}(x,k_T)
+\frac{k_{T\alpha}k_{T\beta}}{4(xp^+)^2}
\Phi_{\bar n11}^{\llbracket \gamma^\alpha\gamma^+\Gamma\gamma^+\gamma^\beta \rrbracket}(x,k_T)\Big],
\end{eqnarray}
where the factor $2k^+$ is the Jacobian for $\delta(k^2)$ and all scalar products are  four-dimensional. The same expression is valid for $\widetilde{\Psi}_n$ after the replacement of $n\leftrightarrow \bar n$.

The twist-two part of a general TMD distribution is related to the matrix element of the free massless parton. In fact, this statement is correct for leading-twist (twist-two) distributions of any kind. For example, in the case of PDF the equivalent derivation can be found in ref.\cite{Balitsky:1987bk}. 

The hadron tensor with inclusion of all KPCs is given by eqn.(\ref{sumKPC:1}) with replacement $\widetilde{\Psi}\to [\widetilde{\Psi}]_{\text{tw-2}}$. For the particular elements of the Dirac basis one finds
\begin{eqnarray}\label{tw2:1}
\Big[\widetilde{\Psi}_{\bar n}^{[1]}(y)\Big]_{\text{tw-2}}&=&0,
\\\label{tw2:5}
\Big[\widetilde{\Psi}_{\bar n}^{[\gamma^5]}(y)\Big]_{\text{tw-2}}&=&0,
\\\label{tw2:v}
\Big[\widetilde{\Psi}_{\bar n}^{[\gamma^\mu]}(y)\Big]_{\text{tw-2}}&=&2p^+
\int dx d^4 k \,\delta(k^+-xp^+)\delta(k^2)e^{i(ky)} k^\mu \Phi_{\bar n 11}^{[\gamma^+]}(x,k_T),
\\\label{tw2:a}
\Big[\widetilde{\Psi}_{\bar n}^{[\gamma^\mu\gamma^5]}(y)\Big]_{\text{tw-2}}&=&2p^+
\int dx d^4 k \,\delta(k^+-xp^+)\delta(k^2)e^{i(ky)} k^\mu \Phi_{\bar n 11}^{[\gamma^+\gamma^5]}(x,k_T),
\\\label{tw2:t}
\Big[\widetilde{\Psi}_{\bar n}^{[i\sigma^{\mu\nu}\gamma^5]}(y)\Big]_{\text{tw-2}}&=&2p^+
\int dx d^4 k \,\delta(k^+-xp^+)\delta(k^2)e^{i(ky)} \Big(
\\\nn &&
g_T^{\mu\alpha}k^\nu-g_{T}^{\nu\alpha}k^\mu 
+\frac{k^\mu n^\nu-n^\mu k^\nu}{k^+}k_T^\alpha
\Big) \Phi_{\bar n 11}^{[i\sigma^{\alpha+}\gamma^5]}(x,k_T).
\end{eqnarray}
One can see that the TMD correlators are divergence-free and satisfy Laplace equation
\begin{eqnarray}
&&\frac{\partial^2}{\partial y^\mu \partial y_\mu }\Big[\widetilde{\Psi}_{\bar n}^{[\Gamma]}(y)\Big]_{\text{tw-2}}=0,
\\
&&
\frac{\partial}{\partial y^\mu}\Big[\widetilde{\Psi}_{\bar n}^{[\gamma^\mu]}(y)\Big]_{\text{tw-2}}
=
\frac{\partial}{\partial y^\mu}\Big[\widetilde{\Psi}_{\bar n}^{[\gamma^\mu\gamma^5]}(y)\Big]_{\text{tw-2}}
=
\frac{\partial}{\partial y^\mu}\Big[\widetilde{\Psi}_{\bar n}^{[i\sigma^{\mu\nu}\gamma^5]}(y)\Big]_{\text{tw-2}}
=0.
\end{eqnarray}
These equations are identical to the equations for twist-two part of collinear operators \cite{Balitsky:1987bk, Balitsky:1989jb, Geyer:1999uq}.

\subsection{Hadron tensor with summed KPCs}
\label{sec:W-summed}

Substituting expressions (\ref{tw2:1} -- \ref{tw2:t}) into (\ref{sumKPC:1}) one obtains the hadron tensor with resummed KPCs. The general expression reads
\begin{eqnarray}\label{KPC-main}
W_{\text{KPC}}^{\mu\nu}&=&-\frac{4p_1^+p_2^-}{N_c}
C_0\(\frac{Q^2}{\mu^2}\)\int d\xi_1d\xi_2 \int d^4k_1 d^4k_2 
\,\delta^4(q-k_1-k_2)
\\\nn &&
\delta(k_1^+-\xi_1p_1^+)
\delta(k_2^--\xi_2p_2^-)
\delta(k_1^2)
\delta(k_2^2)
\Bigg\{
\\\nn &&
\((k_1k_2)g^{\mu\nu}-k_1^\mu k_2^\nu-k_2^\mu k_1^\nu\)\(
\Phi_{\bar n11}^{[\gamma^+]}\Phi_{n11}^{[\gamma^-]}+
\Phi_{\bar n11}^{[\gamma^+\gamma^5]}\Phi_{n11}^{[\gamma^-\gamma^5]}\)
\\\nn &&
+i\epsilon^{\mu\nu \alpha\beta}k_1^\alpha k_2^\beta \(
\Phi_{\bar n11}^{[\gamma^+]}\Phi_{n11}^{[\gamma^-\gamma^5]}-\Phi_{\bar n11}^{[\gamma^+\gamma^5]}\Phi_{n11}^{[\gamma^-]}\)
+t^{\mu\nu}_{\alpha\beta}
\Phi_{\bar n11}^{[i\sigma^{\alpha+}\gamma^5]}
\Phi_{n11}^{[i\sigma^{\beta-}\gamma^5]}\Bigg\},
\end{eqnarray}
where
\begin{eqnarray}\nn
t^{\mu\nu}_{\alpha\beta}=\frac{\Tr[\gamma^\mu \sigma^{\gamma\rho}\gamma^{\nu}\sigma^{\sigma\delta}]}{16}\(
g_T^{\sigma\alpha}k_1^\delta-g_{T}^{\delta\alpha}k_1^\sigma 
+\frac{k_1^\sigma n^\delta-n^\delta k_1^\sigma}{k_1^+}k_T^{\alpha}\)
\(g_T^{\gamma\beta}k_2^\rho-g_{T}^{\rho\beta}k_2^\gamma 
+\frac{k_2^\gamma \bar n^\rho-\bar n^\rho k^\gamma}{k_2^-}k_T^{\beta}\)
\end{eqnarray}
In this expression, the product of two distributions should be understood as following
\begin{eqnarray}
\Phi_{\bar n}^{[\Gamma_1]}\Phi_{n}^{[\Gamma_2]}&=&
\Phi_{\bar n,q}^{[\Gamma_1]}(\xi_1,k_{1T};\mu^2,Q^2)\Phi_{n,\bar q}^{[\Gamma_2]}(\xi_2,k_{2T};\mu^2,Q^2)
\\\nn &&
+\Phi_{\bar n,\bar q}^{[\Gamma_1]}(\xi_1,k_{1T};\mu^2,Q^2)\Phi_{n,q}^{[\Gamma_2]}(\xi_2,k_{2T};\mu^2,Q^2),
\end{eqnarray}
where in the second term the quark and anti-quark TMDPDFs are exchanged. The collinear momentum fractions are denoted by $\xi$ in order to distinguish them from the ordinary Drell-Yan momentum fractions $x$. Note, that in eqn.(\ref{KPC-main}) all indices and momenta are four-dimensional. The argument of the coefficient function is restored to $Q^2$. It is due to the $\delta$-functions in the integrand of eqn.(\ref{KPC-main}), which allows to rewrite $X$ (\ref{def:X}) as $X=2(k_1k_2) =(k_1+k_2)^2=Q^2$.

The expression (\ref{KPC-main}) has the same hard-coefficient function as the LP TMD factorization, and the TMDPDFs are the usual TMDPDFs obeying the standard evolution equations. The main limitation of eqn.(\ref{KPC-main}) is that the rapidity scales are taken in the symmetric point, $\zeta=\bar \zeta=Q^2$. It is necessary in order to remove the terms proportional to $\partial_\mu \mathcal{D}\ln(\bar\zeta/\zeta)$. Otherwise the summation of KPCs is not possible. However, if required, these terms could be restored infinitesimally. Practically, the restriction $\zeta=\bar \zeta=Q^2$ is not essential, since all phenomenological applications are done at the symmetric point.

The expression (\ref{KPC-main}) is the TMD factorization with resummed KPCs and the main result of this work. The dropped terms either contain TMD distributions of higher-twist, either proportional to $q_T/Q$. Being expanded at large-$Q$ $W^{\mu\nu}_{\text{KPC}}$ reproduces the ordinary fixed-power TMD factorization power-by-power.

The hadron tensor $W^{\mu\nu}_{\text{KPC}}$ is exactly transverse to the vector $q^\mu$ (\ref{qW=0}). It is also frame-invariant, up to the corrections which come due to the selection direction of the Wilson lines. The later follows from the the initial assumption on the momenta counting rules for partons. These counting rules also receive power corrections and modify the definition of TMD distributions.  The resulting factorization theorem does not depend on the definition of vectors $n$ and $\bar n$. The inclusion of these corrections goes beyond the present work, and will be performed in the future.

\section{KPC for the unpolarized Drell-Yan reaction}
\label{sec:structure-general-main}

To exemplify the general structure derived above, let me consider the case of the spherically-symmetric contribution to the unpolarized Drell-Yan. In this case, the only\footnote{
Another pair of TMDPDFs that contributes to the unpolarized Drell-Yan hadronic tensor is the pair of Boer-Mulder functions $h_1^\perp$. Generally, all angular structure functions contain contributions proportional to $f_1f_1$ and $h_1^\perp h_1^\perp$, see f.i. refs.\cite{Arnold:2008kf, Balitsky:2021fer}. I have checked explicitly that the spherically-symmetric structure function with KPCs does not contain contribution $\sim h_1^\perp h_1^\perp$.
} contributing TMDPDFs are unpolarized TMDPDFs $f_1$. They are defined as
\begin{eqnarray}
\Phi_{11}^{[\gamma^+]}(x,k_T)&=& f_1(x,k_T)+...~,
\end{eqnarray}
where dots indicate the omitted Sivers function. In this case the series of KPCs (\ref{KPC:general1}) reads
\begin{eqnarray}\label{KPC:f1f1}
W^{\mu\nu}_{f_1f_1}&=&
\frac{p^+_1p^-_2}{N_c}\int d^2k_{1T}d^2k_{2T}\delta^{(2)}(q_T-k_1-k_2) \int dx d\tilde x  \delta(q^+-xp_1^+)\delta(q^--\tilde xp_2^-) 
\\\nn &&
C_0\(\frac{X}{\mu^2}\)\sum_{n,m=0}^\infty \frac{(\partial_{\tilde x})^{n}(\partial_{x})^{m}}{n!m!}\frac{(k_{1T}^2)^n (k_{2T}^2)^m}{(2p^+_1p^-_2)^{n+m}x^n\tilde x^m}\Bigg\{
\\\nn 
&&
-g_T^{\mu\nu}+
\frac{n^\nu k_{1T}^\mu +n^\mu k_{1T}^\nu}{xp_1^+}+\frac{\bar n^\nu k_{2T}^\mu+\bar n^\mu k_{2T}^\nu}{\tilde xp_2^-}
-\frac{g^{\mu\nu}(k_1k_2)_T-k_{1T}^\mu k_{2T}^\nu-k_{1T}^\nu k_{2T}^\mu}{x\tilde xp^+_1p^-_2}
\\\nn
&&
-\frac{\bar n^\mu \bar n^\nu k_{2T}^2}{(\tilde xp_1^-)^2}
-\frac{n^\mu n^\nu k_{1T}^2}{(xp_2^+)^2}
-\frac{\bar n^\mu k_{1T}^\nu+\bar n^\nu k_{1T}^\mu}{2xp_1^+(\tilde xp_2^-)^2}k_{2T}^2
-\frac{n^\mu k_{2T}^\nu+n^\nu k_{2T}^\mu}{2(xp_1^+)^2\tilde xp_2^-}k_{1T}^2
\\\nn 
&&
-g_T^{\mu\nu}\frac{k_{1T}^2k_{2T}^2}{(2x\tilde xp^+_1p^-_2)^2}\Bigg\}f_1(x,k_{1T};\mu,X)f_1(\tilde x,k_{2T};\mu,X),
\end{eqnarray}
where $X$ is given in eqn.(\ref{def:X}). The summed expression (\ref{KPC-main}) turns to
\begin{eqnarray}\label{KPC-main-f1f1}
W_{\text{KPC}f_1f_1}^{\mu\nu}&=&-\frac{4p_1^+p_2^-}{N_c}
C_0\(\frac{Q^2}{\mu^2}\)\int d\xi_1d\xi_2 \int d^4k_1 d^4k_2 
\delta^4(q-k_1-k_2)
\\\nn &&
\delta(k_1^+-\xi_1p_1^+)
\delta(k_2^--\xi_2p_2^-)
\delta(k_1^2)
\delta(k_2^2)
\\\nn &&
\((k_1k_2)g^{\mu\nu}-k_1^\mu k_2^\nu-k_2^\mu k_1^\nu\)
f_1(\xi_1,k_{1T};\mu,Q^2)f_1(\xi_2,k_{2T};\mu,Q^2).
\end{eqnarray}
The same expression could be obtained computing the Drell-Yan reaction with free massless quarks produced by TMD distributions. In other words, this result exactly reproduces the naive parton model. However, it is not naive, because it is known which terms of factorization series were dropped and thus, it could be systematically improved.

For the examples discussed in this section, it is instructive to have the first three terms of the standard power expansion in the explicit form. Let me denote the terms of expansion as
\begin{eqnarray}
W^{\mu\nu}_{f_1f_1}=W_{0,f_{1}f_1}^{\mu\nu}+W_{1,f_{1}f_1}^{\mu\nu}+W_{2,f_{1}f_1}^{\mu\nu}+...~,
\end{eqnarray}
where $W_n\sim Q^{-n}$ in power counting. Explicitly, these terms are
\begin{eqnarray}\label{LP:unpol-example-momentum}
W_{0,f_1f_1}^{\mu\nu}&=&\frac{-g_T^{\mu\nu}}{N_c}
\int d^2k_{1T} d^2k_{2T} \delta^{(2)}(q_T-k_{1T}-k_{2T})C_0
f_{1}(x_1,k_{1T})f_{1}(x_2,k_{2T}),
\\
\label{NLP:unpol-example-momentum}
W_{1,f_1f_1}^{\mu\nu}&=&
\frac{1}{N_c}\int d^2k_{1T} d^2k_{2T} \delta^{(2)}(q_T-k_{1T}-k_{2T})C_0 
\\\nn && \times 
\Big(\frac{\bar n^\mu k_{2T}^\nu+k_{2T}^\mu \bar n^\nu}{q^-}
+\frac{n^\mu k_{1T}^\nu+k_{1T}^\mu n^\nu}{q^+}\Big)f_{1}(x_1,k_{1T}) f_{1}(x_2,k_{2T})
,
\\\label{NNLP:unpol-example-momentum}
W_{2,f_1f_1}^{\mu\nu}&=&
\frac{1}{N_c}\int d^2k_{1T} d^2k_{2T} \delta^{(2)}(q_T-k_{1T}-k_{2T})C_0 
\\\nn && 
\times 
\Bigg[-\frac{g^{\mu\nu}(k_1\cdot k_2)_T -k^\mu_{1T}k^\nu_{2T}-k^\mu_{2T}k^\nu_{1T}}{q^+q^-}
-\frac{n^\mu n^\nu}{(q^+)^2}k_{1T}^2
-\frac{\bar n^\mu \bar n^\nu}{(q^-)^2}k_{2T}^2
\\\nn &&\qquad
+\frac{g_T^{\mu\nu}}{2q^+q^-}\(x_2k_{1T}^2 \frac{\partial}{\partial x_2}
+x_1k_{2T}^2 \frac{\partial}{\partial x_1}\)
\Bigg]f_{1}(x_1,k_{1T}) f_{1}(x_2,k_{2T}),
\end{eqnarray}
where $x_1=q^+/p_1^+$ and $x_2=q^-/p_2^-$. The rapidity-scaling arguments for the LP expression are $f_1(x_1,k_{1T};\mu,\zeta)f_1(x_2,k_{2T};\mu,\bar \zeta)$, and for the NLP they are $f_1(x_1,k_{1T};\mu,2q^+q^-)f_1(x_2,k_{2T};\mu,2q^+q^-)$. The argument for the coefficient function at LP and NLP is $2q^+q^-/\mu^2$. For the (pure) NNLP term one could not specify rapidity scales and the argument of the coefficient function without introducing higher-power terms as it is discussed above.

In the position space expressions for the first three power are
\begin{eqnarray}\label{LP:unpol-example-position}
W_{0,f_1f_1}^{\mu\nu}&=&
\frac{-g_T^{\mu\nu}}{N_c}\int \frac{d^2b}{(2\pi)^2} \, e^{-i(q_Tb)}C_0 
\widetilde{f}_{1}(x_1,b)\widetilde{f}_{1}(x_2,b),
\\
\label{NLP:unpol-example-position}
W_{1,f_1f_1}^{\mu\nu}&=&
\frac{1}{N_c}\int \frac{d^2b}{(2\pi)^2} \, e^{-i(q_Tb)}C_0 
\\\nn && \times 
\Big(-\frac{i\bar n^\mu }{q^-}\widetilde{f}_{1}(x_1,b) D^\nu\widetilde{f}_{1}(x_2,b)
-\frac{in^\mu}{q^+}D^\nu\widetilde{f}_{1}(x_1,b) \widetilde{f}_{1}(x_2,b)\Big)
,
\\
\label{NNLP:unpol-example-position}
W_{2,f_1f_1}^{\mu\nu}&=&
\frac{1}{N_c}\int \frac{d^2b}{(2\pi)^2} \, e^{-i(q_Tb)}C_0 
\bigg[\frac{g^{\mu\nu} D_\alpha \widetilde{f}_{1} D^\alpha \widetilde{f}_{1}-D^\mu \widetilde{f}_{1} D^\nu \widetilde{f}_{1}-D^\nu \widetilde{f}_{1}D^\mu \widetilde{f}_{1}}{q^+q^-}
\\\nn && 
+\frac{n^\mu n^\nu}{(q^+)^2}D^2\widetilde{f}_{1}\widetilde{f}_{1}
+\frac{\bar n^\mu \bar n^\nu}{(q^-)^2}\widetilde{f}_{1}D^2\widetilde{f}_{1}
+\frac{g_T^{\mu\nu}}{2q^+q^-}\(x_1\frac{\partial \widetilde{f}_{1}}{\partial x_1}D^2\widetilde{f}_{1}+x_2D^2\widetilde{f}_{1}\frac{\partial \widetilde{f}_{1}}{\partial x_2}\)
\bigg],
\end{eqnarray}
where arguments for TMDPDFs are omitted completely for the NNLP case. The operator $D$ is the ``long'' derivative defined in eqn.(\ref{def:D}). In contrast to the momentum-space expressions (\ref{LP:unpol-example-momentum} -- \ref{NNLP:unpol-example-momentum}), the position space expressions are boost-invariant (\ref{boost-invariance}), and the rapidity-scaling arguments are $\zeta$ and $\bar \zeta$.

The expressions (\ref{LP:unpol-example-position} - \ref{NNLP:unpol-example-position}) can be compared to the literature. Naturally, the NLP part agrees with the one computed in ref.\cite{Vladimirov:2021hdn}. Another computation was performed in refs.\cite{Balitsky:2020jzt, Balitsky:2021fer} using the small-x approximation. These results can be directly compared to eqn.(\ref{KPC:f1f1}). I have found that the LP and NLP parts (\ref{LP:unpol-example-position}, \ref{NLP:unpol-example-position}) agree with ref.\cite{Balitsky:2020jzt}, but the NNLP part disagrees. The same holds for the double Boer-Mulders term, which is not presented here. The hadron tensor derived in refs.\cite{Balitsky:2020jzt, Balitsky:2021fer} is transverse exactly at NNLP, while (\ref{KPC:f1f1}) is transverse in the sum of all powers. Possibly, it is due to the differences in the definition of twist-four terms.

\subsection{Restoration of EM gauge invariance}
\label{sec:EM-example}

The EM gauge-invariance (or electric-charge conservation) imposes
\begin{eqnarray}\label{dJ=0}
\partial_\mu J^\mu=0,
\end{eqnarray}
which on the level hadronic tensor turns to
\begin{eqnarray}\label{qW=0}
q_\mu W^{\mu\nu}= W^{\nu\mu}q_\mu=0.
\end{eqnarray}
This statement is exact in QCD. Nonetheless, EM gauge-invariance is violated at each power of TMD factorization. It is a consequence of the fact that the relation (\ref{qW=0}) does not have a definite power counting, since it contains a sum of $q^\pm$ and $q_T$. The EM gauge-invariance is restored by the power corrections. Moreover , it is restored for each terms with unique nonperturbative content by a the series of KPCs.

Let me demonstrate the restoration of EM gauge invariance using the unpolarized DY example in the momentum space
\footnote{The same exercise can be done without assumptions on $\zeta$'s in the position space. In this case, the integrand of $\widetilde{W}^{\mu\nu}$ should be acted by the operator $\hat q^\mu=\bar n^\mu q^++n^\mu q^--i\frac{\partial}{\partial b_\mu}$. The results of this section are reproduced also for the terms that contain derivatives of the Collins-Soper kernel.}
. Contracting the LP term (\ref{LP:unpol-example-momentum}) with $q^\mu$, one obtains
\begin{eqnarray}\label{EM:qw0}
q_\mu W_{0,f_1f_1}^{\mu\nu}&=&\frac{-1}{N_c}
\int d^2k_{1T} d^2k_{2T} \delta^{(2)}(q_T-k_{1T}-k_{2T})C_0 
\\\nn && \times 
(k_{1T}^{\nu}+k_{2T}^{\nu}) f_{1}(x_1,k_{1T})f_{1}(x_2,k_{2T})
\neq 0.
\end{eqnarray}
So, the EM gauge-invariance is violated, but the violation goes beyond the LP approximation, because the left-hand-side(LHS) of the formula is $\sim Q^1$ while the RHS is $\sim Q^0$. The inclusion of the NLP term (\ref{NLP:unpol-example-momentum}) leads to
\begin{eqnarray}\label{EM:qw0+1}
&&q_\mu (W_{0,f_1f_1}^{\mu\nu}+W_{1,f_1f_1}^{\mu\nu})=\frac{1}{N_c}
\int d^2k_{1T} d^2k_{2T} \delta^{(2)}(q_T-k_{1T}-k_{2T})C_0 
\\\nn &&\qquad\qquad \times 
\Big[\(\frac{\bar n^\nu (k_2q)_T}{q^-}+\frac{n^\nu (k_1q)_T}{q^+}\) f_{1}(x_1,k_{1T})f_{1}(x_2,k_{2T})\Big]\neq 0.
\end{eqnarray}
The RHS expression (\ref{EM:qw0+1}) is of order $\sim Q^{-1}$, i.e. NNLP in comparison to the LHS. The LP violation term (\ref{EM:qw0}) is canceled by the NLP violation term, but leaves an NNLP remnant. The twist-three contributions are not presented here, but it could be checked using expression in ref.\cite{Vladimirov:2021hdn, Balitsky:2021fer, Ebert:2021jhy} that they also result into $\sim Q^{-1}$ terms. The NNLP remnant in eqn.(\ref{EM:qw0+1}) is then canceled by NNLP (\ref{NNLP:unpol-example-momentum}) term:
\begin{eqnarray}\label{EM:qw0+1+2}
&&q_\mu (W_{0,f_1f_1}^{\mu\nu}+W_{1,f_1f_1}^{\mu\nu}+W_{2,f_1f_1}^{\mu\nu})=\frac{1}{N_c}
\int d^2k_{1T} d^2k_{2T} \delta^{(2)}(q_T-k_{1T}-k_{2T})C_0 
\\\nn && \times
\[\frac{k_{1T}^\nu k_{2T}^2+k_{2T}^\nu k_{1T}^2}{q^+q^-}
-\frac{q_T^{\nu}}{2q^+q^-}\(x_2k_{1T}^2 \frac{\partial}{\partial x_2}
+x_1k_{2T}^2 \frac{\partial}{\partial x_1}\)
\] f_{1}(x_1,k_{1T})f_{1}(x_2,k_{2T})
\neq 0.
\end{eqnarray}
Here, the RHS is of order $\sim Q^{-2}$, i.e. N$^3$LP in comparison to LHS. This N$^3$LP violation term is to be canceled by the corresponding part of $q_\mu W_{3}^{\mu\nu}$, and so on. 

In this way, EM gauge invariance is restored only in the sum of all powers of TMD factorization, and violated at any fixed power.  A truncation of the series at power $n$ results into the violation of EM gauge invariance at order $1/Q^{n+1}$. This mechanism is standard for the factorization theorems with inhomogeneous counting of $q^\mu$ components. Another very well studied example is DVCS, see refs.\cite{Belitsky:2001hz, Belitsky:2010jw, Braun:2011zr, Braun:2011dg} for detailed discussion.

Importantly, the restoration of EM gauge invariance takes place independently for the terms proportional to twist-two TMDPDFs, and twist-three TMDPDFs. This is due to the proper definition of the TMD-twist. With the present definition TMD distributions of different twists do not mix with each other, and are entirely independent nonperturbative functions. Each independent combination of TMDPDFs forms its own series of KPCs that restores the EM gauge invariance.

The complete series (\ref{KPC:f1f1}) is exactly EM-gauge invariant. The check is straightforward, one should only take into account that $q^\pm$ turns to $xp$ by $\delta$-function and does not commute with the derivative. All terms of expression (\ref{KPC:f1f1}) participate in the restoration, which is in contrast to the small-x-based derivation made in ref.\cite{Balitsky:2021fer}, where NNLP term already completes the gauge-invariant sequence. The summed expression (\ref{KPC-main}) is also EM-gauge invariant since $q^\mu=k_1^\mu+k_2^\mu$ and
\begin{eqnarray}
q^\mu\((k_1k_2)g^{\mu\nu}-k_1^\mu k_2^\nu-k_2^\mu k_1^\nu\)
=
-k_1^2 k_2^\nu-k_2^2k_1^\nu=0,
\end{eqnarray}
because $k_{1,2}^2=0$ due to the $\delta$-functions. The part $\sim t^{\mu\nu}$ in (\ref{KPC-main}) is also transverse.

The cancellation between successive terms takes place only if their hard coefficient functions are identical. This has been checked up in sec.~\ref{sec:argument-Q} explicitly at NLO, but the gauge-invariance guaranties the equivalence at all perturbative orders. The EM-gauge invariance does not determine the coefficient function of other genuine contributions, since they have independent nonperturbative content.

\subsection{Restoration of frame invariance}
\label{sec:frame-example}

Another important symmetry that is violated by the factorization approach is the Lorenz invariance. In the factorization theorems approach it is often called the frame-invariance or the reparameterization invariance. It has been intensively studied for the case of collinear factorization for DVCS, see refs.\cite{Braun:2011dg, Braun:2014sta}, and also in the soft-collinear/heavy-quark effective fields theories, see refs.\cite{Luke:1992cs, Manohar:2002fd, Marcantonini:2008qn}. Meanwhile, I do not know any dedicated discussion for the TMD factorization case.

The frame- or reparameterization-invariance is based on the fact that the field-mode separation depends on the directions $n$ and $\bar n$, which, in turn, are defined only approximately. In other words, the counting rules (\ref{counting-rules}) can be modified by power corrections, if they preserve the leading counting. The light-cone vector $n^\mu$, which defines the collinear direction, could be turned $n^\mu\to n'^\mu$ by a power-suppressed amount such that $n'^2=0$, and the factorization theorem remains the same. If there are several collinear directions the transformation should preserve the relation between them. This is an obvious statement in the deep-inelastic scattering. In other cases the frame-invariance is rather involved, and usually is violated by the LP term and restored by power corrections similarly to EM gauge invariance. 

Let me inspect the implication of frame-invariance for the TMD factorization approach. There are two collinear directions given by $n^\mu$ and $\bar n^\mu$, which satisfy $(n\bar n)=1$. It is straightforward to see that there are two possible transformations $n\to n'$, that preserve $n'^2=\bar n'^2=0$ and $(n'\bar n')=1$. They are
\begin{equation}\label{def:frame-transform}
\text{I:~} \left\{\begin{array}{l}
\Ds n^\mu\to n'^\mu=n^\mu+\frac{\Delta^\mu}{q^-}-\frac{\Delta^2}{2(q^-)^2}\bar n^\mu,
\\
\Ds \bar n^\mu\to \bar n'^\mu=\bar n^\mu,
\end{array}\right.
\qquad
\text{II:~} \left\{\begin{array}{l}
\Ds n^\mu\to n'^\mu=n^\mu,
\\
\Ds \bar n^\mu\to \bar n'^\mu=\bar n^\mu+\frac{\overline{\Delta}^\mu}{q^+}-\frac{\overline{\Delta}^2}{2(q^+)^2}n^\mu,
\end{array}\right.
\end{equation}
where $\Delta$ and $\overline \Delta$ are transverse vectors in the original frame, i.e. $(\Delta n)=(\Delta \bar n)=0$. The vectors have the counting $\Delta \sim Q^0$, and could be treated as infinitesimal parameters. Note, that one cannot transform $n$ and $\bar n$ simultaneously, because in this case, the equation $(n\bar n)=1$ requires $\Delta \sim Q^1$, which is not a small transformation. Nonetheless, transformations I and II can be applied successively, simulating a simultaneous rotation of $n$ and $\bar n$. It is important to mention that the vectors $p^\mu_1$ and $p_2^\mu$ are external vectors, and thus they are not transformed under (\ref{def:frame-transform}). The definition (\ref{def:p1p2}) is not modified by $\Delta$.

The transformations I and II are symmetric to each other. Therefore, the result valid for one of them is automatically valid for another. Thus, in the rest of the section,  I consider only the transformation I. 

On the level of factorized expression transformations (\ref{def:frame-transform}) change the definition of TMDPDFs, since $(k_Tn')\neq0$. Therefore, for the comparison of momentum space integrands (similar analysis can be done for position space integrands), one should compensate the redefinition of $k$'s, such that the definition of $f_1(x,k_T)$ remains the same. Recalling that TMDPDF is obtained by Fourier transforms (\ref{def:TMD-momentum}) one finds that the same definitions can be achieved by the contra-rotation
\begin{equation}\label{def:frame-transform-k}
k^\mu_{1T}\to {k'}_{1T}^{\mu}=k^\mu_{1T},
\qquad
k^\mu_{2T}\to {k'}_{2T}^{\mu}=k_{2T}^\mu-\Delta^\mu-((k_2\Delta)-\Delta^2)\frac{\bar n^\mu}{q^-}.
\end{equation}
Here, $(k_{1T}\Delta)=0$ is imposed, in order to preserve the definition of TMDPDF $f_1(x_1,k_{1T})$. The variables $x$ are also impacted by the rotation because they are defined via $q^\pm$. One finds that
\begin{eqnarray}\label{def:frame-transform-x}
x_1\to x'_1=x_1\(1+\frac{2 (k_2\Delta)-\Delta^2}{2 q^+q^-}\),\qquad x_2\to x_2'=x_2,
\end{eqnarray}
where it is used that $(q\Delta)=(k_2\Delta)$, due to the delta-function. So, the simultaneous transformations (\ref{def:frame-transform}) (part I), (\ref{def:frame-transform-k}) and (\ref{def:frame-transform-x}) should keep the integrand of the factorized hadronic tensor invariant.

Applying the transformation to the LP term (\ref{LP:unpol-example-momentum}) one finds
\begin{eqnarray}
W_{0,f_1f_1}^{\mu\nu}&\to& W_{0,f_1f_1}^{\mu\nu}+\frac{1}{N_c}
\int d^2k_{1T} d^2k_{2T} \delta^{(2)}(q_T-k_{1T}-k_{2T})C_0 
\Big\{
\\\nn &&
\frac{\bar n^\mu \Delta^\nu+\Delta^\mu n^\nu}{q^-}+...\Big\}f_1(x_1,k_{1T})f_1(x_2,k_{2T}),
\end{eqnarray}
where the dots denote the higher power terms. The first term in brackets is $\sim Q^{-1}$ i.e. NLP. So, the LP term is frame-invariant up to NLP corrections. Adding the NLP contribution (\ref{NLP:unpol-example-momentum}) one obtains
\begin{eqnarray}
W_{0,f_1f_1}^{\mu\nu}&+&W_{1,f_1f_1}^{\mu\nu}\to W_{0,f_1f_1}^{\mu\nu}+W_{1,f_1f_1}^{\mu\nu}+\frac{1}{N_c}
\int d^2k_{1T} d^2k_{2T} \delta^{(2)}(q_T-k_{1T}-k_{2T})C_0 
\Big\{
\\\nn &&
\(\frac{\Delta^\mu k_{1T}^\nu+k_{1T}^\mu \Delta^\nu}{q^+q^-}+\frac{\bar n^\mu \bar n^\nu(\Delta^2-2(k_2\Delta))}{(q^-)^2}\)f_1f_1
-x_1g_T^{\mu\nu}\frac{2(k_2\Delta)-\Delta^2}{2q^+q^-}\frac{\partial f_1}{\partial x_1}f_1+...\Big\}.
\end{eqnarray}
The arguments of TMDPDFs are omitted but the relative order of TMDPDFs is preserved. Again the violation term starts with $\sim Q^{-2}$ and is NNLP. The derivative term is obtained from the expansion of $x_1'$. Finally, one can check that
\begin{eqnarray}
W_0^{\mu\nu}&+&W_1^{\mu\nu}+W_2^{\mu\nu}\to W_0^{\mu\nu}+W_1^{\mu\nu}+W_2^{\mu\nu}+\mathcal{O}(\text{N$^3$LP}),
\end{eqnarray}
where $\mathcal{O}(\text{N$^3$LP})$ is a long expression. Importantly, that the frame-invariance involves the derivative-of-TMDPDF terms, which do not participate in the check of gauge-invariance at NNLP order. Therefore, the series of KPCs cannot be split into sub-series that are gauge and frame invariant. Alike the gauge-invariance, the frame-invariance is not complete at any given power-order. Moreover, at each power order it generates the infinite series of derivative corrections, due to the Taylor expansion of $f_1(x'_1,k_{1T})$, which altogether restore the frame-invariance of the full series. The frame invariance holds for each nonperturbative sector, i.e. separately for the power series involving  twist-two TMDPDFs, separately for the power series involving twist-three TMDPDFs, etc.

The expression with summed KPCs (\ref{KPC-main-f1f1}) is invariant under the transformations (\ref{def:frame-transform}) since it does not depend vectors $n$ and $\bar n$. Here, I recall that the external momenta $p_1^\mu$, $p_2^\mu$ are not transformed, and thus $\delta$-functions with arguments $(\xi p^\pm-k^\pm)$ are not modified.

\subsection{Estimation of numerical importance}
\label{sec:numerics}

The factorization theorem with summed KPCs is valid in the traditional regime of TMD phenomenology, $Q\gg q_T$ and $Q\gg\Lambda$ but does not have restriction $Q\gg k_T$. In fact, the restriction $Q\gg k_T$ was ignored in all phenomenological studies, assuming that $k_T\sim \Lambda$. The assumption $k_T\sim \Lambda$ is not entirely correct because $k_T$ is an integration variable. In this section, I test the impact of inclusion of KPCs. For it, I compare LP and KPC-summed cross-sections of the unpolarized Drell-Yan (only $\gamma$-channel for simplicity).

The cross-section for the Drell-Yan reaction is computed from the hadronic tensor by
\begin{eqnarray}
\frac{d\sigma}{dQ^2dyd\vec q_T^2}=\frac{2\alpha_{\text{em}}}{sQ^4}\sum_{q} e_q^2\int d\Omega \, L_{\mu\nu}W^{\mu\nu},
\end{eqnarray}
where $\vec q_T^2=-q_T^2>0$, $\alpha_{\text{em}}$ is the QED coupling, and $e_q$ is the charge of the quark $q$. The leptonic tensor is
\begin{eqnarray}
L^{\mu\nu}=4(l^\mu l'^{\nu}+l'^\mu l^\nu-g^{\mu\nu}(ll')),
\end{eqnarray}
where $l$ and $l'$ are leptons' momenta and $l+l'=q$.

Substituting the hadronic tensor from eqn.(\ref{LP:unpol-example-momentum}) one obtains the cross-section in the LP TMD factorization
\begin{eqnarray}\label{num:LP}
\frac{d\sigma}{dQ^2dydq_T^2}\Big|_{\text{LP}}&=&\frac{(2\pi)^2\alpha_{\text{em}}}{3N_csQ^2}\(1+\frac{\vec q_T^2}{2Q^2}\)
\sum_{q} e_q^2 C_0\(\frac{2q^+q^-}{\mu^2}\)
\\\nn && \times
\int d^2\vec k_{1T}d^2\vec k_{2T} \delta^{(2)}(\vec q_T-\vec k_{1T}-\vec k_{2T})f_{1q}(x_1,k_{1T};\mu,\zeta)f_{1\bar q}(x_2,k_{2T};\mu,\bar \zeta),
\end{eqnarray}
where 
\begin{eqnarray}\label{num:xi}
x_1=\frac{q^+}{p_1^+}=\frac{Q e^y}{\sqrt{s}}\sqrt{1+\frac{\vec q_T^2}{Q^2}},
\qquad
x_2=\frac{q^-}{p_2^-}=\frac{Q e^{-y}}{\sqrt{s}}\sqrt{1+\frac{\vec q_T^2}{Q^2}},
\end{eqnarray}
The correction $\vec q_T^2/2Q^2$ in the common factor is the result of convolution of the leptonic tensor with $g_T^{\mu\nu}$. The inhomogeneity of this factor in power-counting is due fact that the leptonic tensor is exact (i.e. it contains all powers) while the hadronic tensor is pure LP. Exactly this expression is usually used for the phenomenology of unpolarized TMD distributions, see e.g.\cite{Scimemi:2017etj, Scimemi:2019cmh, Moos:2023yfa}.

\begin{figure}
\begin{center}
\includegraphics[width=0.4\textwidth]{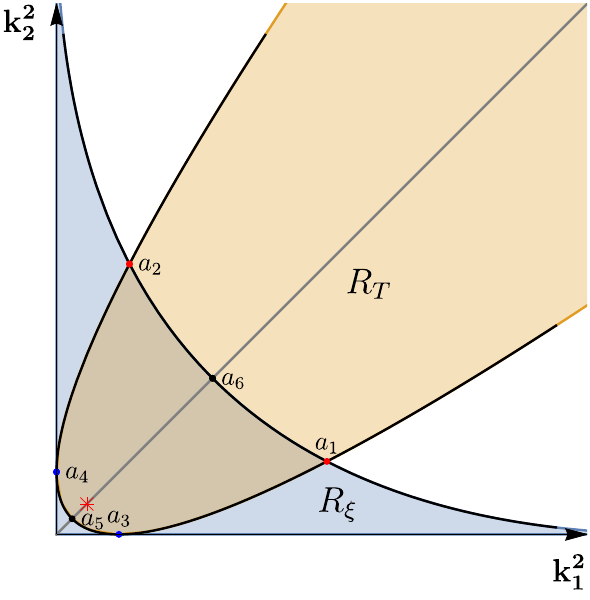}
~
\includegraphics[width=0.45\textwidth]{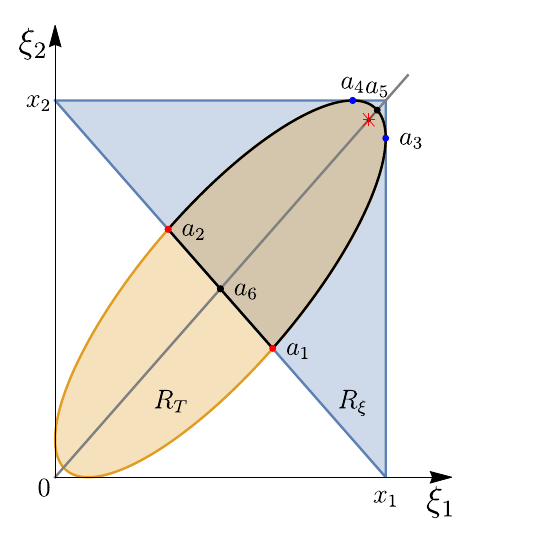}
\end{center}
\caption{\label{fig:phasespace} (left) The region of integration for variables $\vec k_{1T}^2$ and $\vec k_{2T}^2$ in the TMD factorization formulas (\ref{num:LP}) and (\ref{num:KPC}). (right) The region of values of parameters $\xi_{1,2}$ (\ref{num:xi}) covered during the integration over $\vec k_{1T}^2$ and $\vec k_{2T}^2$. The points $a_i$ at both plots corresponds to the same values of $\vec k_{1T}^2$ and $\vec k_{2T}^2$. }
\end{figure}

The integral over transverse momenta $k_{1T}$ and $k_{2T}$ is convenient to present as the integral over $\vec k_{1T}^2$ and $\vec k_{2T}^2$. The delta-function restricts the values of these variables as
\begin{eqnarray}
R_T: \qquad \vec k_{1T}^2>0,\qquad \vec k_{1T}^2+\vec q_T^2-\sqrt{\vec k_{qT}^2\vec q_T^2}<\vec k_{2T}^2<\vec k_{1T}^2+\vec q_T^2+\sqrt{\vec k_{qT}^2\vec q_T^2}.
\end{eqnarray}
The region $R_T$ is shown in the fig.\ref{fig:phasespace}(left) by yellow color. Notably, the region spans to infinite values of $\vec k_T^2$'s, despite initial assumption of $Q$ being the largest scale.

The cross-section with resummed KPC is obtained from eqn.(\ref{KPC-main}) and reads
\begin{eqnarray}\label{num:KPC}
\frac{d\sigma}{dQ^2dydq_T^2}\Big|_{\text{KPC}}&=&\frac{(2\pi)^2\alpha_{\text{em}}}{3N_cs}\sum_{q} e_q^2 C_0\(\frac{Q^2}{\mu^2}\)
\int d^4k_{1}d^4k_{2} \int d\xi_1 d\xi_2 \delta^{(4)}(q-k_{1}-k_{2})
\\\nn && \nn
\delta(k_1^2)\delta(k_2^2)\delta(k_1^+-\xi_1p_1^+)\delta(k_2^--\xi_2p_2^-)
f_{1q}(\xi_1,k_{1T};\mu,Q^2)f_{1\bar q}(\xi_2,k_{2T};\mu,Q^2).
\end{eqnarray}
It is straightforward to check that in the limit $q^\pm\to\infty$ the expression (\ref{num:KPC}) reproduces (\ref{num:LP}) (up to the power-suppressed term in the common factor). Note that to make this comparison one should commute the limit $Q\to\infty$ and the integration operation. In other words, one should assume that $\vec k^2_T\ll Q^2$, despite the integral range of $\vec k_T^2$'s is infinite. 

The values collinear momentum-fractions of TMDPDFs are not fixed in the summed formula (\ref{num:KPC}), but integrated in a particular range. It could be anticipated apriory, because the reparametrization invariance modifies the value of $x$'s (\ref{def:frame-transform-x}), and thus any fixed choice is not frame-invariant. The $\delta$-functions express $\xi$'s as
\begin{eqnarray}\label{def:xi}
\xi_1&=&\frac{x_1}{2}\(1+\frac{\vec k_{1T}^2}{\tau^2}-\frac{\vec k_{2T}^2}{\tau^2}+\frac{\sqrt{\lambda(\vec k_{1T}^2,\vec k_{2T}^2,\tau^2)}}{\tau^2}\),
\\\nn
\xi_2&=&\frac{x_2}{2}\(1-\frac{\vec k_{1T}^2}{\tau^2}+\frac{\vec k_{2T}^2}{\tau^2}+\frac{\sqrt{\lambda(\vec k_{1T}^2,\vec k_{2T}^2,\tau^2)}}{\tau^2}\),
\end{eqnarray}
where $\tau^2=2q^+q^-=Q^2+\vec q_T^2$, and $\lambda(a,b,c)=a^2+b^2+c^2-2ab-2ac-2bc$ is the kinematic function. The values of $\xi_{1,2}$ are restricted $0<\xi_{1,2}<1$ (due to the support properties of TMDPDF), which constraints the integration region of $\vec k_{1T}^2$ and $\vec k_{2T}^2$ to
\begin{eqnarray}
R_\xi: \qquad 0<\vec k_{1T}^2<\tau^2,\qquad 0<\vec k_{2T}^2<\tau+\vec k_{1T}^2-2\sqrt{\vec k_{1T}^2 \tau^2}.
\end{eqnarray}
The region $R_\xi$ is shown in the fig.\ref{fig:phasespace}(left) by blue color. The variables $\xi$ (\ref{def:xi}) can be seen as a kind-of-Nachmann variables \cite{Nachtmann:1973mr, Georgi:1976ve} for the TMD factorization theorem, which correct for the convolution integral for non-zero transverse momenta of partons.

\begin{figure}
\begin{center}
\includegraphics[width=0.4\textwidth]{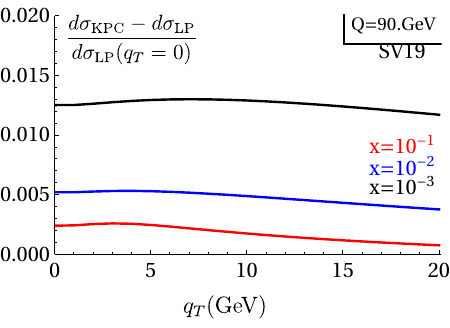}
~
\includegraphics[width=0.4\textwidth]{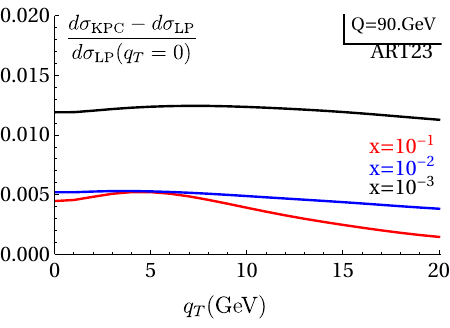}
\\
\includegraphics[width=0.4\textwidth]{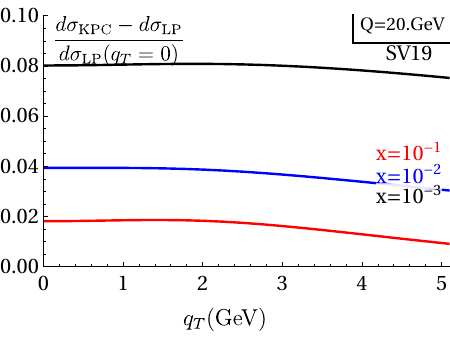}
~
\includegraphics[width=0.4\textwidth]{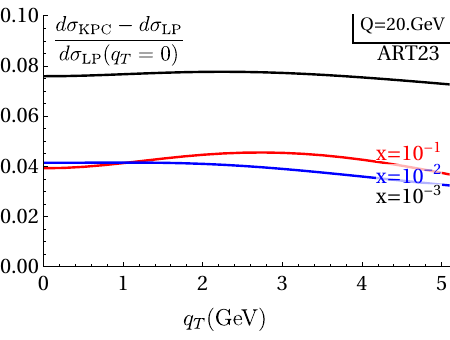}
\end{center}
\caption{\label{fig:vsqT} The difference of KPC-summed and LP cross-sections relative to the $q_T=0$ LP cross-section for the Drell-Yan reaction at different $q_T$ and different $x$. The plot in left(right) panels are based on SV19 \cite{Scimemi:2019cmh} (ART23 \cite{Moos:2023yfa}) extraction of TMDPDF. Upper row is for $Q=90$GeV, and lower row is for $Q=20$GeV. Lines of different color corresponds to different values of $x$ (at $y=0$).}
\end{figure}

In full the integration region for (\ref{num:KPC}) is $R_\xi \cap R_T$. It is compact. The maximum values of $\vec k_T^2$ are $(\sqrt{\tau^2}+\sqrt{\vec q_T^2})^2/4$ ($= Q^2/4$ at $\vec q_T^2=0$) which are reached at the points $a_{1,2}$ in fig.\ref{fig:phasespace}. The arguments $\xi_{1,2}$ belongs to the region shown in fig.\ref{fig:phasespace}(right). This region does not include the LP values of collinear momentum-fraction $x_{1,2}$ (\ref{num:xi}), but it includes the values $\sqrt{Q^2/s}e^{\pm y}$, which are often used in the phenomenology and is shown by the red star in fig.\ref{fig:phasespace}.

\begin{figure}
\begin{center}
\includegraphics[width=0.45\textwidth]{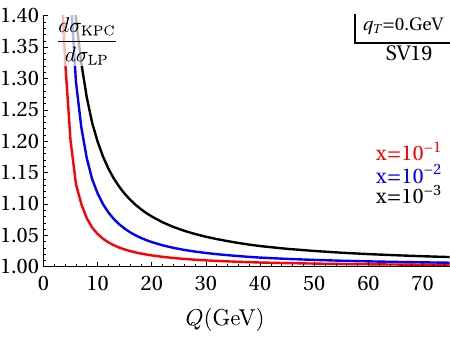}
~
\includegraphics[width=0.45\textwidth]{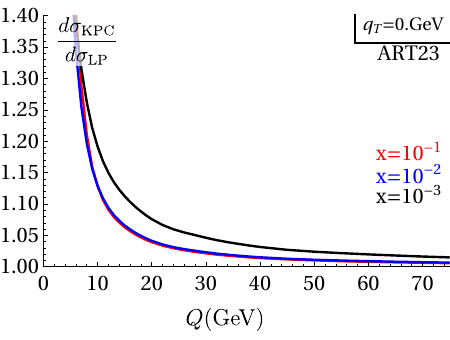}
\end{center}
\caption{\label{fig:vsQ}  The ratio of KPC-summed to LP cross-sections for the Drell-Yan reaction at $q_T=0$ at different $Q$ and different $x$. The plot in left(right) panel is based on SV19 \cite{Scimemi:2019cmh} (ART23 \cite{Moos:2023yfa}) extraction of TMDPDF. Lines of different color corresponds to different values of $x$ (at $y=0$).}
\end{figure}

In figs.~\ref{fig:vsqT} and \ref{fig:vsQ}, the comparison of KPC-summed (\ref{num:KPC}) and LP (\ref{num:LP}) cross-sections is shown. For the comparison the extractions SV19 \cite{Scimemi:2019cmh} (presented in the left panels) and ART23 \cite{Moos:2023yfa} (presented in the right panels) were used. The perturbative orders are the same as used in the extractions. They are N$^3$LL for SV19, and N$^4$LL for ART23. The evolution of distributions is performed in the position space, then the TMDPDFs are Fourier-transformed and substituted into (\ref{num:KPC}, \ref{num:LP}). The results for different extraction have generally the same size but somewhat different behavior at $x>10^{-2}$.

The inclusion of KPCs results into an almost flat increase of the cross-section as a function of $q_T$, see fig.\ref{fig:vsqT}. The size of the shift grows at smaller $x$'s and $Q$. In other words, the inclusion of KPCs does not significantly modify the shape of the LP prediction, but significantly changes the normalization. At the typical kinematic of the LHC (aka Z-boson production) the size of corrections is of the order of $1$\%. It grows to $\sim 30-40$\% at $Q\sim 10$GeV. Using the presented curves one can estimate the effective size of KPCs. Their order of magnitude is approximately described by
\begin{eqnarray}
\frac{d\sigma|_{\text{KPC}}}{d\sigma|_{\text{LP}}}\sim 1+ \frac{(2\text{GeV})^2}{x^{0.4} Q^2}.
\end{eqnarray}
This approximation is obtained by a two-parameter fit of SV19 and ART23 curves.

It is interesting to mention that the modern TMD phenomenology faces problems with the description of normalization of cross-section, while perfectly describes the shape. For LHC energies the issue is of order of few percents, while for low-energy fix-target data ($Q\sim 5-20$GeV) is tenths of percents \cite{Bertone:2019nxa, Bacchetta:2019sam, Vladimirov:2019bfa}. The situation is worse for $Q\sim 2-4$GeV which are typical for Semi-Inclusive Deep-Inelastic Scattering (SIDIS). In this case the problem with normalization is of order of factor 2-3 \cite{Bacchetta:2022awv}. Therefore, one could hope that inclusion of KPCs into the phenomenology will resolve this important problem.

\section{Conclusion}

The power corrections in the TMD factorization theorem provide a rich field for investigation. As described in the introduction, there are four conceptual types of power corrections: $q_T/Q$, $k_T/Q$, $\Lambda/Q$, and target-mass corrections. These corrections have distinct origins and characteristics. Importantly, different power corrections are significant in different kinematic regimes, which justifies considering them independently. In this work, I have derived the series of kinematic power corrections (KPCs) or $k_T/Q$-corrections that follow the leading power (LP) term. These corrections play a special role for the factorization theorem and are essential for its consistency because KPCs are responsible for restoring electromagnetic (EM) gauge invariance (charge conservation) and frame invariance, which are broken by the LP approximation. In that sense, the series of KPCs is the entailed part of the LP factorization theorem.

The KPCs that follow the LP term possess a distinctive mark: they only involve TMD distributions of twist-two. This feature allows for their straightforward extraction from the generic power expansion of operators. The resulting series is presented in equation (\ref{KPC:general1}). This series is summed in a simple expression (\ref{KPC-main}), which is the main result of the work. The computation is done for the Drell-Yan process but can be generalized for other processes without conceptual problems. 

One of the most significant features of KPCs is that they must obey the factorization theorem, even if other types of power corrections may violate it. This statement follows from the fact that KPCs are responsible for restoring the fundamental properties of the LP term, namely charge conservation and frame invariance. Consequently, the coefficient function for KPCs remains identical to the LP coefficient function. In sec.~\ref{sec:argument-Q}, I have explicitly verified this statement at the next-to-leading order (NLO) and demonstrated the restoration of the argument of the LP coefficient function to $q^2$. The latter is a non-trivial result as it indicates the impossibility of a strict power expansion beyond NLP.

The final formula for the cross-section is almost trivial -- it tells that the cross-section can be computed with free massless quarks similar to a naive parton picture. However, with the present derivation, it attains a different status. It obeys the factorization theorem, and it is clear which part of the power expansion is included and which is neglected. The excluded terms consist of TMD distributions of twist-three and higher, as well as powers of $q_T/Q$

It is important to note that the derivation of the series of KPCs is contingent upon the definition of TMD-twist and higher twist distributions. The current derivation is based on the approach outlined in ref.~\cite{Vladimirov:2021hdn}, which leads to a consistent result. However, if an alternative definition is proposed, the series of KPCs could differ. In such a case, some of the present higher twist terms would be absorbed into the KPC series. In particular, the difference in the treatment of higher twist terms may describe the discrepancy between the present computation and ref.~\cite{Balitsky:2020jzt}.

Incorporating KPCs into TMD phenomenology is essential, as they play a crucial role in restoring the consistency of the formalism. Moreover, all perturbative ingredients remain unchanged since KPCs adhere to the LP factorization theorem. Therefore, including KPCs in phenomenological studies can be done without encountering conceptual difficulties while maintaining the achieved level of accuracy (currently at the N$^4$LL order \cite{Moos:2023yfa}). The summed formula is valid in the same kinematic range of application as ordinary TMD factorization. Estimations made in sec.~\ref{sec:numerics} demonstrate that including KPCs results in an almost constant increment of the cross-section. The magnitude of this correction depends on $Q$ and $x$. For typical LHC kinematics, the correction is around 1\%, while at $Q\sim 4-5$ GeV, the correction can reach 100\%. Interestingly, the deficiency in normalization for the TMD factorization at low energies has been reported by multiple groups. One could expect that these problems will be resolved with the inclusion of KPCs.

\acknowledgments
I thank Volodia Braun, Ian Balitky, Oscar del Rio, Simone Rodini, and Ignazio Scimemi for numerous discussions and helpful comments. I also thank Sara Alvarez for critical reading and corrections to the manuscript. This work is funded by the \textit{Atracci\'on de Talento Investigador} program of the Comunidad de Madrid (Spain) No. 2020-T1/TIC-20204, and is also supported by the Spanish Ministry grant PID2019-106080GB-C21.

\appendix

\bibliography{bibFILE}

\end{document}